%% file: main.tex
\documentclass[sigconf, nonacm=true]{acmart}

\def\BibTeX{{\rm B\kern-.05em{\sc i\kern-.025em b}\kern-.08emT\kern-.1667em\lower.7ex\hbox{E}\kern-.125emX}}

\usepackage{xspace}
\usepackage{threeparttable}
\usepackage{pifont}
\usepackage{booktabs}
\usepackage{multirow}
\usepackage{todonotes}
\usepackage{siunitx}
\usepackage{textcomp}
\usepackage[absolute,showboxes]{textpos}

\DeclareSIUnit\permille{\text{\textperthousand}}
\DeclareSIUnit{\sample}{S}
\DeclareSIUnit\Fahrenheit{\degree F}
\usepackage {tikz}
\newcommand*\circled[1]{\tikz[baseline=(char.base)]{
		\node[shape=circle,draw,inner sep=1pt,fill=orange,text=white] (char) {\bfseries{\textsf{#1}}};}}

\definecolor{dgreen}{rgb}{0.0, 0.5, 0.0}

\newcommand{\sname}{\textsf{Eco}\xspace}
\newcommand{\cmark}{\color{dgreen}\ding{51}}%
\newcommand{\mcmark}{\color{orange}(\ding{51})}%
\newcommand{\xmark}{\color{red}\ding{55}}%
\newcommand{\etal}{\textit{et al.}~}
\newcommand{\eg}{\textit{e.g.,}~}
\newcommand{\ie}{\textit{i.e.,}~}

\newcommand{\one}{({\em i})\xspace}
\newcommand{\two}{({\em ii})\xspace}
\newcommand{\three}{({\em iii})\xspace}
\newcommand{\four}{({\em iv})\xspace}

\let\orgautoref\autoref
\renewcommand{\autoref}
{\def\sectionautorefname{Section}%
	\def\subsectionautorefname{Section}%
	\def\subsubsectionautorefname{Subsection}%
	\orgautoref}

\newenvironment{sitemize}{%
\begin{list}{$\bullet$}{
	\setlength{\itemsep}{0.0cm}%
	\setlength{\leftmargin}{2.4em}%
	\setlength{\topsep}{0cm}%
	\setlength{\parsep}{0mm}}%
}{\end{list}}

\setcopyright{none}
\begin{document}

\title[Eco: A HW-SW Co-Design for In Situ Power Measurement on Low-end IoT Systems]{Eco: A Hardware-Software Co-Design for In Situ Power Measurement on Low-end IoT Systems}

\author{Michel Rottleuthner}
\affiliation{%
	\institution{HAW Hamburg}
}
\email{michel.rottleuthner@haw-hamburg.de}

\author{Thomas C. Schmidt}
\affiliation{%
	\institution{HAW Hamburg}
}
\email{t.schmidt@haw-hamburg.de}

\author{Matthias W{\"a}hlisch}
\affiliation{%
	\institution{FU Berlin}
}
\email{m.waehlisch@fu-berlin.de}

\setlength{\TPHorizModule}{\paperwidth}
\setlength{\TPVertModule}{\paperheight}
\TPMargin{5pt}
\begin{textblock}{0.8}(0.1,0.02)
	\noindent
	\footnotesize
	If you cite this paper, please use the ENSsys reference:
	Michel Rottleuthner, Thomas C. Schmidt, and Matthias W\"ahlisch. 2019. Eco:
	A Hardware-Software Co-Design for In Situ Power Measurement on Low-
	end IoT Systems. In \emph{Proc. of the 7th Int. Workshop on Energy Harvesting \& Energy-Neutral Sensing Systems (ENSsys '19)}, ACM, 2019.
\end{textblock}

\input{abstract}

\input{category_keywords}
\maketitle
\input{introduction}
\input{ps_and_related_work}
\input{implementation}
\input{evaluation}

\input{deployment}
\input{conclusions}
\input{acks}
\balance

\bibliographystyle{ACM-Reference-Format}
\bibliography{iot,own,manet,rfcs}

\appendix
\end{document}

%% file: abstract.tex
\begin{abstract}
Energy-constrained sensor nodes can adaptively optimize their energy consumption if a continuous measurement exists.
This is of particular importance in scenarios of high dynamics such as energy harvesting or adaptive task scheduling. However, self-measuring of power consumption at reasonable cost and complexity is unavailable as a generic system service. 
In this paper, we present \sname, a hardware-software co-design enabling generic energy management on IoT nodes.
\sname is tailored to devices with limited resources and thus targets most of the upcoming IoT scenarios.
The proposed measurement module combines commodity components with a common system interfaces to achieve easy, flexible integration with various hardware platforms and the RIOT IoT operating system.
We thoroughly evaluate and compare accuracy and overhead. Our findings indicate  that our commodity design competes well with highly optimized solutions, while being significantly more versatile. 	
We employ \sname for energy management on RIOT and validate its readiness for deployment in a five-week field trial integrated with energy harvesting.
\end{abstract}

%% file: category_keywords.tex
%
\ccsdesc[500]{Computer systems organization~Embedded systems}
\ccsdesc[500]{Hardware~Energy metering}
\keywords{Energy harvesting, power measurement, IoT}

%% file: introduction.tex
\section{Introduction}
\label{sec:intro}

Energy is a scarce resource in the  constrained Internet of Things (IoT) of independently powered  nodes,  many of which need to balance tasks with the amount of energy available for sustaining operation over their targeted lifetime.

Software development paradigms for these IoT systems recently shift away from static bare-metal code towards flexible, platform independent applications that offer reusable functionality.
Nonetheless, building energy-aware systems is still a task that requires developers to deeply engage with many low-level, platform-specific details.
Partially this arises because energy availability and power consumption are much more dynamic than system properties such as processing and memory resources.
Energy levels fluctuate, in particular on nodes that harvest energy from environmental sources.
This makes it crucial for a system to track its power input and consumption, so that runtime tasks can adapt to energy availability quick and effectively.

Energy harvesting describes a process where the output of typically weak power sources is collected and stored until a sufficient amount of energy is available to perform a desired task.
In the context of low-end IoT networks, energy harvesting allows for building independently powered systems with the potential to sustain perpetual operation, thereby increasing performance and reducing maintenance cost \cite{basm-ehwts-16}.
Because of the volatile nature of common energy sources, energy harvesting demands for management mechanisms to effectively use energy and avoid system outages.

Tracking, controlling, and optimizing energy consumption depends on many aspects and  quickly becomes complex.
Therefore, simulations combined with prior external measurements are often applied to estimate power flows.
However, for systems that are subject to varying operational or environmental conditions, it is often infeasible to rely entirely on \emph{a priori} lab testing.
Simplified models and error-prone parameter estimators imply further inaccuracies. Accurate information about actual deployment conditions are required instead.
For those reasons \emph{in situ} measurement is a key strategy to obtain real-world data.
For this a generic off-the-shelf solution is desired that is portable, reusable, and covers various settings of software and hardware components.

\begin{figure}[h]
	\includegraphics[width=0.5\textwidth]{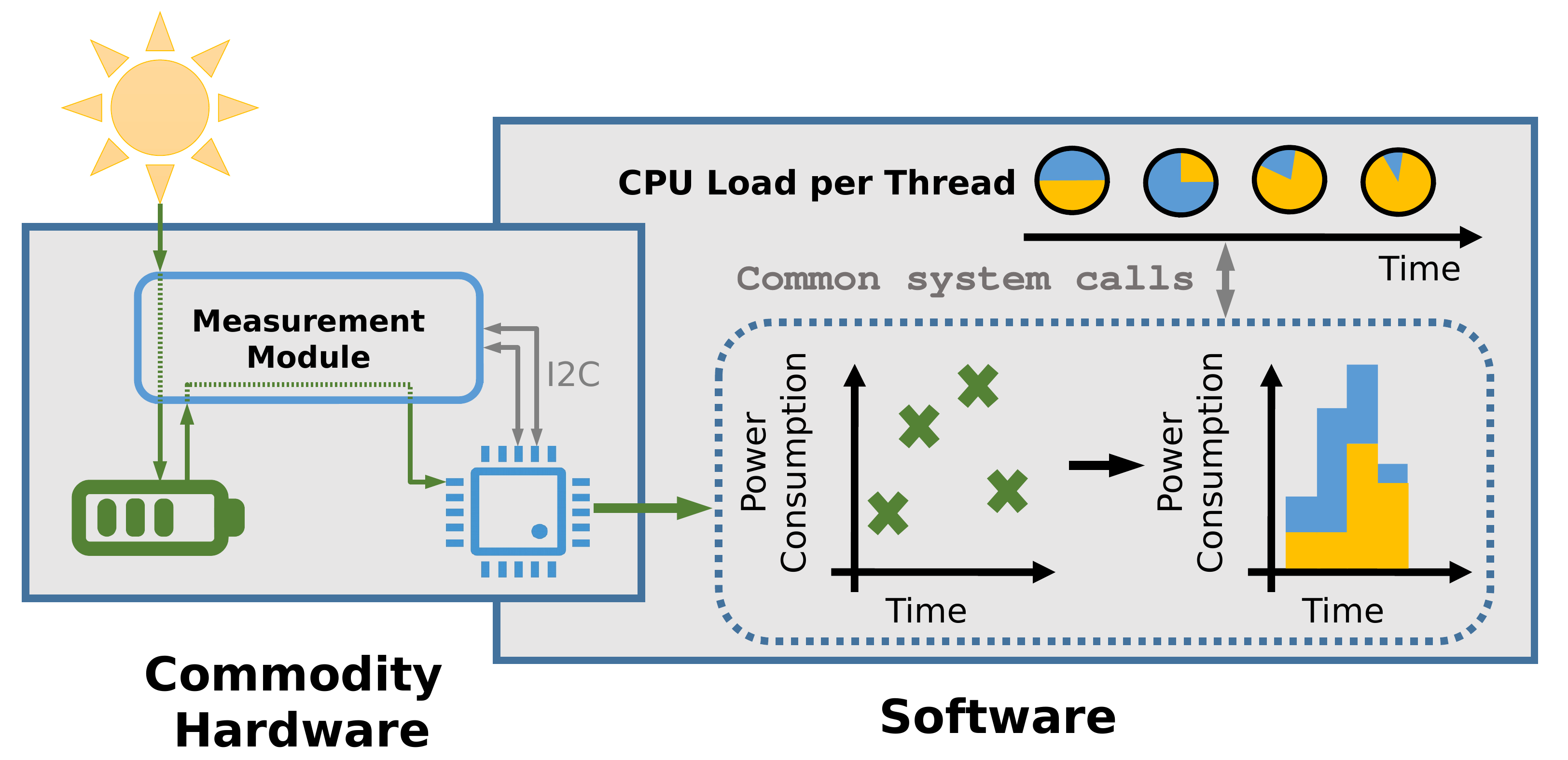}
	\caption{\sname measures energy in situ by monitoring power sources and consumers.}
	\label{fig:eco-overview}
\end{figure}

In this paper, we introduce \sname, a hardware-software co-design enabling generic energy management on low-end commodity IoT devices (see \autoref{fig:eco-overview}).
As a generally applicable approach, \sname is based on two key principles.
First, it  connects a simple external measurement module via the standard  I$^2$C interface which is widely available on all targeted microcontrollers.
Only the control pins require connecting and the power supply line needs to be routed through this module.
Second, it uses platform-independent software interfaces that allow for an easy change of microcontroller boards without any need of software adaptation.

In detail, we contribute:
\begin{sitemize}
  \item A novel key component enabling energy management on autonomous IoT nodes that are battery-driven, have few kilobytes of memory, and MHz of CPU power.
  \item An  implementation of \sname in the IoT operating system RIOT that fosters deployment, and allows for online energy evaluation of individual tasks and threads.
  \item An evaluation of \sname including its cross-platform validation and the detailed quantification of errors and overheads.
  \item An extensive real-world deployment within an energy harvesting system as a comprehensive proof of concept of the vertical integration of hardware and software in the field.
\end{sitemize} 

\noindent \sname outperforms the current state-of-the-art.
The closest related results from the literature~\cite{dfpc-emfas-08} focus on the plain measurement problem without higher level system integration.
This specialized, component-free approach has lower current measurement accuracy (them 10\% vs. \sname 1\%), lower resolution (them \SI{10}{\micro\ampere} vs. \sname \SI{1.25}{\micro\ampere}), but lower energy overhead (them $\approx$0.05\% vs. \sname $\approx$~1\%).

It consists of a few commodity components, only, and allows for measuring current and voltage over adjustable ranges and with flexible sampling rates.
Hence, it is easy to reproduce and can be attached to a large variety of IoT boards.

The remainder of this paper continues with discussing the energy management problem together with related work.
The design of our measurement module in hardware and software is presented (\S~\ref{sec:implementation}), followed by a thorough evaluation (\S~\ref{sec:evaluation}).
We report on our deployment experiment and the lessons learned thereof (\S~\ref{sec:deployment}).
Finally, we conclude with an outlook in \autoref{sec:conclusions}.

%% file: ps_and_related_work.tex
\section{The Problems of Dynamic Energy \\ Management and Related Work}
\label{sec:problem}
Any approach to managing energy on constrained devices relies on accurate knowledge of the actual conditions.
Overestimating the available energy may quickly put a node out of service, underrating energy may hinder its operational utility.
In a variable, dynamic setting it becomes important to obtain a thorough understanding of the various energy flows, including the consumption of different application tasks and potentially available inflows.

The key problem of dynamic energy management arises from the difficulty to determine power consumption timely, with sufficient accuracy, and at tolerable overhead.
Methods for quantifying consumption include theoretical or \mbox{(semi-)}empirical estimation, simulation of simplistic models, as well as external or online self-measurements.
Previous work on measuring consumption of nodes typically focuses either on local lab setups for very accurate monitoring or very light-weight in situ evaluation.

\subsection{Simulation and Estimation of Power \\ Consumption}
Software approaches range from simulation of large scale sensor networks \cite{shcaw-spcls-04} to estimations based on offline reference measurements \cite{doth-soees-07, deft-pnppl-11}.
The prediction accuracy of the power consumption and battery lifetime is significantly affected by its level of abstraction \cite{lwg-appcs-05}.
Neglecting low-level system events such as scheduling and timer related interrupt handling can lead to substantial errors because they often account for a relevant part of the power consumption.

Consumption simulation can be performed from a generic high level perspective \cite{arcsj-sipen} down to estimating CPU cycles \cite{shcaw-spcls-04} or even on the instruction level \cite{tlp-assns-05}.
The full control over relevant system parameters eases isolated analysis of individual aspects.
Downsides relate to the inaccurate reproduction of reality due to environmental changes, varying hardware tolerances, and many other dynamics.
Even if conditions and hardware are exactly the same, different device instances can exhibit significant variations in their consumption \cite{lmgf-oeeme-13}.
Also, simulations mainly focus on improving \emph{a priori} adjustments, leaving runtime optimizations open to other solutions.

Dunkels \etal \cite{deft-pnppl-11} developed Powertrace, a software solution to allow network level profiling of applications for Contiki~\cite{dgv-clfos-04}.
It feeds static values from offline measurements into a linear power model.
This light-weight software-only solution has many benefits compared to hardware solutions.
Unfortunately, it does not apply to dynamically powered systems such as Energy-Harvesting systems with varying supply voltage.
It is also unsuitable when the individual consumption of the various components are tied together with non-linear dependencies.

Estimating the energy consumption {\em online} allows incorporating more runtime specific criteria from actually tracking the system states \cite{deft-pnppl-11}.
As an example, energy usage caused by packet retransmissions can be accounted with higher accuracy, if the exact number of transmissions is known and considered at runtime.
While this varying size can be easily determined in software, other changes such as the efficiency of various electric components are more challenging to estimate.
Incorporating runtime information on the actual consumption as feedback for the energy management algorithm was already shown to improve application level performance and robustness against uncertainties \cite{gjkz-gmoes-19}.

\subsection{Measuring Power Consumption}
A core problem of measuring the power consumption of IoT nodes comes with the heavily varying power demands required by the different power states of the micro controller unit (MCU).
These can span over five orders of magnitude~\cite{jdcs-mpmem-07}.
Additionally, the measurement itself should have little to no side effects on the observed system. 
Combining both introduces further complexity.

\paragraph{External Observers}
Many systems were designed as external observers to record detailed behavior of the sensor node.
The underlying architectures range from Linux-capable systems based on single board computers which cannot be deployed in the field \cite{sglll-mvehi-2017,kzssk-ipcms-17,lfzws-ftdst-13,lt-tacft-17,sk-mdtdw-13}, to small add-on boards that are equipped with additional hardware like an MCU \cite{rs-efsnp-10, zx-nhfnp-13} and in some cases also field programmable gate arrays (FPGA) \cite{rs-efsnp-10, thbr-ahsan-11, zsfs-ebgbs-11}.
Energy-Harvesting is specifically targeted by custom mobile measurement platforms to even allow observation of multi source harvesters \cite{sglll-mvehi-2017} and by providing tools to record and replay harvesting conditions~\cite{zsfs-ebgbs-11}.

With Rocketlogger, Sigrist \etal \cite{sglll-mvehi-2017} introduced a portable device intended to provide a balance between top notch lab equipment and mobile measurement.
The platform provides four voltage and two current channels besides the option to interface digital sensors for additional environment metering. 
Though, the device is based on a BeagleBone Green that alone consumes \SI{7}{\milli\watt} in deep sleep mode and multiples of \SI{100}{\milli\watt} when active. 
Thus, long term off-grid deployment is out of~scope.

Kazdaridis \etal \cite{kzssk-ipcms-17} use classic shunt metering with two resistors in series. 
For dynamic switching, a load switch bypasses the measured current around the high resistance when the burden voltage becomes too high.
The switch is controlled by an analog high speed comparator.
The measurement module is interfaced over I$^2$C by a BeagleBone Black, which introduces power consumption of \SIrange{8.7}{14.2}{\milli\watt}, still ignoring two additionally needed operational amplifiers.
Continuous measurement without MCU interaction are not possible with the module, because no internal sampling buffer or averaging is available.

FlockLab by Lim \etal \cite{lfzws-ftdst-13} targets distributed tracing and profiling.
Only a single shunt is used and linear regression is utilized for calibration.
Additional focus is laid on precise time stamping with an accuracy of around \SI{50}{\micro\second} to correlate events of different nodes.
They also use hardware with relatively high computational power based on a \SI{624}{\mega\hertz} CPU with \SI{128}{\mega\byte} of RAM, disqualifying it for in-situ usage.

\paragraph{Self-Measurement}

A typical way to keep track of the energy flow on the node itself is to utilize coulomb counters. While the temporal resolution is high enough to assess the state of charge, it is insufficient to attribute energy usage to specific tasks or peripheral hardware.
To overcome this issue, custom hardware for faster sampling is required.

When the measurement function is implemented within the IoT device itself, it is crucial to know the overhead that is associated with the measurement.
This overhead concerns power usage of the measurement circuit, memory usage, and CPU time allocated to performing measurement and calculation.

The power measurement based on voltage to frequency conversion introduced by Jiang \etal \cite{jdcs-mpmem-07} is suited to be directly deployed together with a sensor node. 
The system called SPOT leverages a voltage to frequency based digitization process to overcome a lot of the previously described measurement challenges.
While this solution keeps processing overhead low for infrequent reading, it is high for fast sampling.
Furthermore, the discrete components require around \SI{1.7}{\milli\ampere}, which introduces overhead that conflicts with battery powered nodes in the field~\cite{dfpc-emfas-08}.

An implementation by Dutta \etal \cite{dfpc-emfas-08} simplifies SPOT by essentially eliminating the need for any hardware if the platform is powered by a switching regulator.
With this, a resolution down to the \SI{1}{\micro\joule} scale is achieved while staying within a \SI{\pm20}{\percent} error margin.
By using an MCU-internal counter peripheral a read latency down to \SI{15}{\micro\second} can be achieved, leading to a power overhead of only \SIrange{0.01}{0.1}{\percent}.
Shortcomings of the solution are related to inherently high manufacturing tolerances for inductors (around \SI{\pm10}{\percent}), the frequency dependent power overhead, and that the voltage is assumed constant instead of being actually measured.
Other approaches either use highly specialized FPGA implementations for the measurement task \cite{aghfg-tphse-18}, or indirectly assess the consumption by measuring voltage changes at the energy storage element \cite{rtr-oease-14}.

\subsection{Attributing Energy to Software}
The software running on a wireless sensor node defines how energy is spent, \eg by issuing a sensing cycle or requesting the hardware to transmit data.
As resources such as CPU, RAM, or hardware peripherals are dynamically shared between a multitude of software components, attributing the exact amount of used energy to the correct software instance is challenging.
Therefore, the main problem in this domain is correlating power consumption with the executed software.

The granularity and constraints for such attribution mechanisms vary a lot depending on the targeted area of use.
For operators of data-centers it may be enough to know how much energy is used per virtual machine and a little overhead is basically negligible.
On mobile devices the constraints are already much tighter and attributing energy to individual applications is crucial, \eg to prevent faulty ones from draining the battery.
Typical low-end IoT devices further tighten the room for overhead and the granularity is scaled down to smaller entities such as threads, specific tasks or even single function calls.

The simplest base metric for power usage correlation is \emph{utilization}, derived from the time a thread occupies the CPU.
A much more accurate measure to assess CPU-utilization and correlated power consumption is based on monitoring CPU-internal performance counters \cite{b-beeap-00}.
With pTop \cite{drs-ppppt-09}, an implementation for desktop-scale devices was shown that uses this information to attribute power consumption to running processes.

Further accuracy improvements to online estimations can be achieved by tracking power states of individual components.
For TinyOS, Kellner~\cite{k-foeat-10} uses a common model to estimate the overall power consumption which is then attributed to different individual TinyDB queries with the help of resource containers.
Fonseca~\etal \cite{fdls-qtene-08} augment the tracking of component power states with real power measurement and activity tracking to allow fine-grained offline analysis of energy usage.

As TinyOS does not provide threads by default, both solutions introduce abstract entities (\ie activities and resource containers) to which the energy use is attributed.
This allows grouping semantically related resource usage and thereby improves its high level visibility to the developer but requires additional instrumentation of the target application.

\begin{table}[t]
\resizebox{\columnwidth}{!}{
	\begin{threeparttable}
		\centering
		\caption{Comparison of related work}
		\vspace{0.25cm}
		\label{tbl:overview_comparison}
		\begin{tabular}{lccccr}
			\toprule
			\multicolumn{1}{c}{\rotatebox{0}{Variant}} & \multicolumn{1}{c}{\rotatebox{80}{In situ\tnote{\textdagger}}} &
			\multicolumn{1}{c}{\rotatebox{80}{Adaptive}} & \multicolumn{1}{c}{\rotatebox{80}{\parbox{40pt}{Physical\\  Measurem.}}} & \multicolumn{1}{c}{\rotatebox{80}{Portability}} & \multicolumn{1}{c}{\rotatebox{0}{Measurement Control}} \\
			\midrule
			PowerTOSSIM \cite{shcaw-spcls-04} & \xmark & \xmark & \xmark  & \xmark & Offline Simulation \\
			\emph{Kazdaridis et al.} \cite{kzssk-ipcms-17} & \xmark & \xmark & \cmark & \cmark & Passive HW Interface \\
			Rocket Logger \cite{sglll-mvehi-2017} & \xmark & \xmark & \cmark & \cmark &  Time-triggered Logging \\
			FlockLab \cite{lfzws-ftdst-13} & \xmark & \xmark & \cmark & \xmark & Time-triggered Logging \\
			AVEKSHA \cite{thbr-ahsan-11} & \xmark & \xmark & \cmark & \xmark &  Time-triggered Logging \\
			Powertrace \cite{deft-pnppl-11} & \cmark & \cmark & \xmark & \cmark & Programmable Tracing \\
			\emph{Kellner} \cite{k-foeat-10} & \cmark & \cmark & \xmark & \cmark &  Container Tracking \\
			SPOT \cite{jdcs-mpmem-07} & \mcmark \tnote{*} & \mcmark & \cmark & \cmark & Passive HW Interface \\
			Nemo \cite{zx-nhfnp-13} & \mcmark \tnote{*} & \mcmark & \cmark & \xmark & Time-triggered Logging \\
			iCount \cite{dfpc-emfas-08} & \cmark & \mcmark & \cmark & \xmark & Passive SW-Interface \\
			Quanto \cite{fdls-qtene-08} & \cmark & \mcmark & \cmark & \xmark &  Event-triggered Logging \\
			\multirow{2}{*}{\sname} & \multirow{2}{*}{\cmark} & \multirow{2}{*}{\cmark} & \multirow{2}{*}{\cmark} & \multirow{2}{*}{\cmark} & Thread Tracking \&  \\
			 & & & & & Programmable Tracing\\
			\bottomrule
		\end{tabular}
	    \begin{tablenotes}\footnotesize
			\item[\textdagger] Referring to usability in \emph{off-grid} in situ deployments (\ie some of the solutions marked as unsuitable may still be usable for \emph{wired} in situ)
			\item[*] depending on sufficient power supply (refer to \autoref{tbl:comparison} for details)
		\end{tablenotes}
	\end{threeparttable}}
\end{table}

\autoref{tbl:overview_comparison} gives a brief qualitative overview on the discussed work.
In the following sections, we introduce a flexible measurement setup that uses readily available parts and is easy to integrate over an I$^2$C bus.
It spans a wide configurable measurement range, provides different sampling rates and is compatible with common IoT Platforms.
Using various configurations the accuracy is verified with reference measurements.
Additionally, the overhead (\ie invasiveness) induced by the measurement action itself and communication with the module is analyzed to show what cost is tied to more fine-grained energy profiling.
The results can be used to choose an appropriate measurement configuration for specific use cases by weighing between tolerable overhead and additional granularity.

%% file: implementation.tex
\section{Integrated Energy Management:\\ Design and Implementation}
\label{sec:implementation}
\begin{figure*}
\begin{minipage}{.44\textwidth}
	\begin{center}
		\begin {tikzpicture}[-latex, semithick, font=\rmfamily\small, minimum size=25pt,align=center, node distance=.28cm, inner sep=4pt,
		state/.style ={ rectangle ,top color =white , bottom color = black!10 , draw,black , text=black},
		modmarker/.style={circle, draw=black!80, line width=0.2mm, minimum size=8pt,
			,align=center, inner sep=2pt, text=white, fill=orange, font=\bfseries\sffamily\large},
		line/.style={draw, -latex},
		lineda/.style={draw, latex-latex},
		notecircle/.style={circle, draw, minimum size=10pt, inner sep=1pt, black!75, font=\sffamily}]
		\node[state] (ESRC) [] {Power Source};
		\node[state, right=0.4cm of ESRC.east] (CV) [] {Converter};
		\node[state, right=0.4cm of CV.east] (M) [] {Measurement};
		\node[state, right=0.4cm of M.east] (ESTR) [] {Energy Storage};
		\node[state, below=0.6cm of M.south] (MCU) [] {\ \ \ MCU\ \ \ };
		\node[state, left=0.4cm of MCU.west] (TX) [] {TX};
		\node[state, right=0.4cm of MCU.east] (S1) [] {Sensors};
		\node[state, below=0.5cm of MCU.south] (STRG) [] {Persistent Data Storage};
		\path [line] (ESRC) -> (CV);
		\path [line] (CV) -> (M);
		\path [lineda] (M) -- (ESTR);
		\path [line] (M) -- (MCU);
		\path [-, black!75, dashed] (MCU) edge (S1);
		\path [-, black!75, dashed] (MCU) edge (TX);
		\path [-, black!75, dashed] (MCU) edge (STRG);

		\node[modmarker] (A) at (MCU.north west)  {A};
		\node[modmarker] (B) at (TX.north west)   {B};
		\node[modmarker] (C) at (STRG.north west) {C};
		\node[modmarker] (D) at (ESTR.north west) {D};
		\node[modmarker] (E) at (CV.north west)   {E};
		\node[modmarker] (F) at (M.north west)    {F};
	\end{tikzpicture}
\end{center}
\caption{Model of energy harvesting wireless sensor node with self-measurement of power consumption}
\label{fig:eh_sys_model}
\end{minipage}
\hspace{1cm}
\begin{minipage}{.49\textwidth}\centering
	\begin{tikzpicture}[
	notecircle/.style={circle, draw=black!80, line width=0.2mm, minimum size=7.5pt,
		inner sep=0pt, text=white, fill=red, font=\bfseries\sffamily\scriptsize},
	modmarker/.style={circle, draw=black!80, line width=0.2mm, minimum size=8pt,
		,align=center, inner sep=2pt, text=white, fill=orange, font=\bfseries\sffamily\large},
	sysiomarker/.style={opacity=0.4, draw=black!80, line width=0.2mm,
		,align=center, inner sep=2pt, text=white, fill=red, font=\bfseries\sffamily\large},
	sensiomarker/.style={opacity=0.4, draw=black!80, line width=0.2mm,
		,align=center, inner sep=2pt, text=white, fill=yellow, font=\bfseries\sffamily\large},
	pvmarker/.style={opacity=0.4, draw=black!80, line width=0.2mm,
		,align=center, inner sep=2pt, text=white, fill=yellow, font=\bfseries\sffamily\large},
	connectortxt/.style={opacity=1,align=center, inner sep=2pt, text=white, font=\bfseries\sffamily\tiny}]
	\centering\node (img) {\includegraphics[width=.72\columnwidth, angle=0]{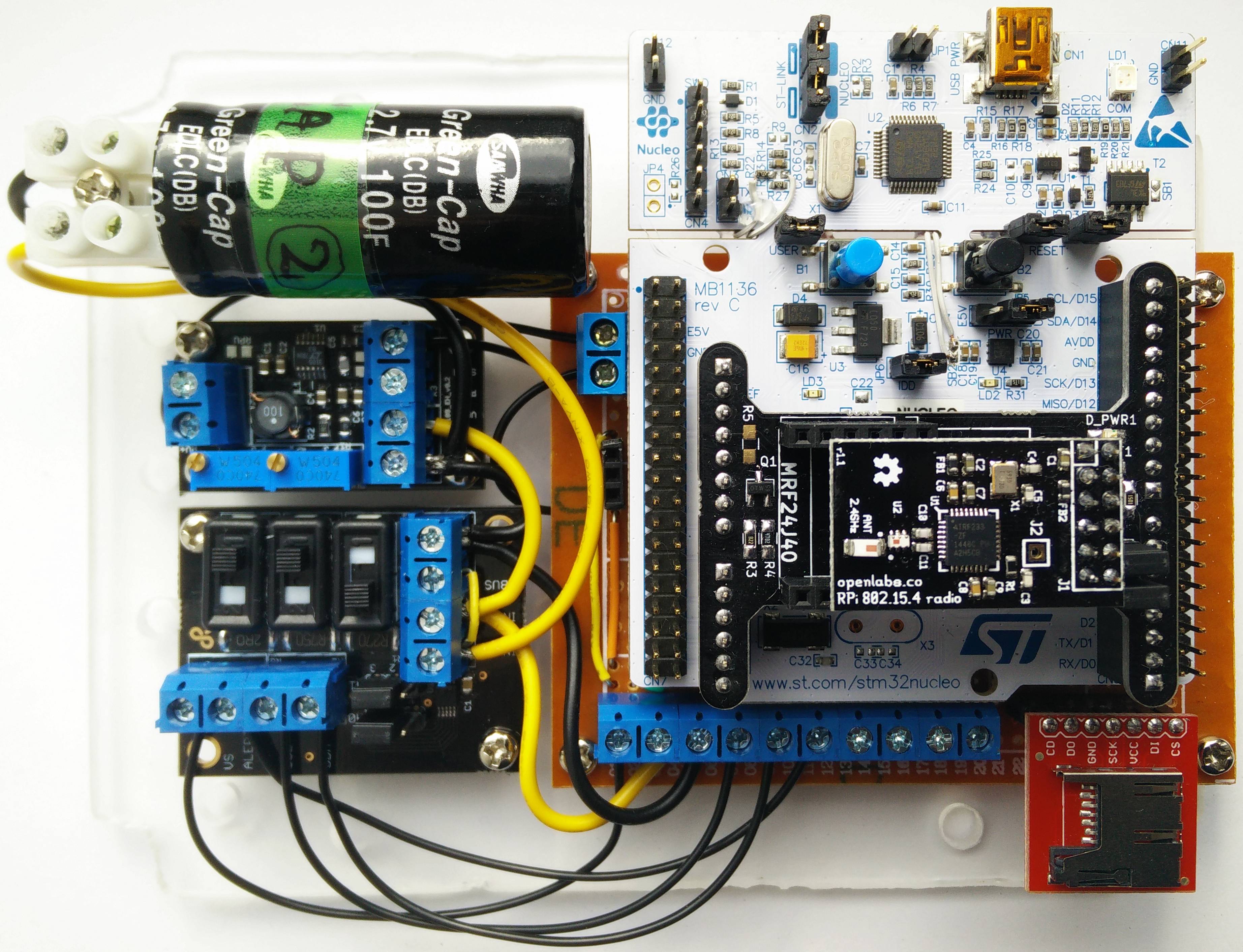}};

	\node[modmarker] (A) at (1.6,0.8)    {A};
	\node[modmarker] (B) at (2,-0.25)    {B};
	\node[modmarker] (C) at (2.6,-1.65)  {C};
	\node[modmarker] (D) at (-1,1.4)     {D};
	\node[modmarker] (E) at (-1.3,0.4)   {E};
	\node[modmarker] (F) at (-1.3,-0.8)  {F};

	\fill[sysiomarker] (-0.15,-1.12) rectangle (1.12,-1.46) node[connectortxt, pos=.5] {Power interface};
	\fill[sensiomarker] (1.13,-1.12) rectangle (1.92,-1.46) node[connectortxt, pos=.5] {Sensor IO};
	\fill[sysiomarker] (-0.23,0.83) rectangle (0.05,0.39);
	\fill[pvmarker] (-1.87,0.58) rectangle (-2.35,0.13) node[connectortxt, pos=.5, rotate=-90] {PV};

	\end{tikzpicture}\vspace*{-16pt}
	\caption{Mounting plate with interconnected modules of the energy-harvesting system}%
	\label{fig:eh_motherboard}%
\end{minipage}
\end{figure*}

The integrated hardware-software co-design is outlined in the following.
First, a flexible modular hardware platform is introduced.
\autoref{fig:eh_sys_model} shows the abstract hardware architecture consisting of a wireless sensor node and a power subsystem.
The power subsystem includes a power source, modules for charging and measurement and an energy storage element.
The node itself is built as a fully modular combination of MCU, radio transceiver, sensors and persistent data storage.

\begin{figure}[b]
	\centering
	\includegraphics[width=0.7\columnwidth, angle=0]{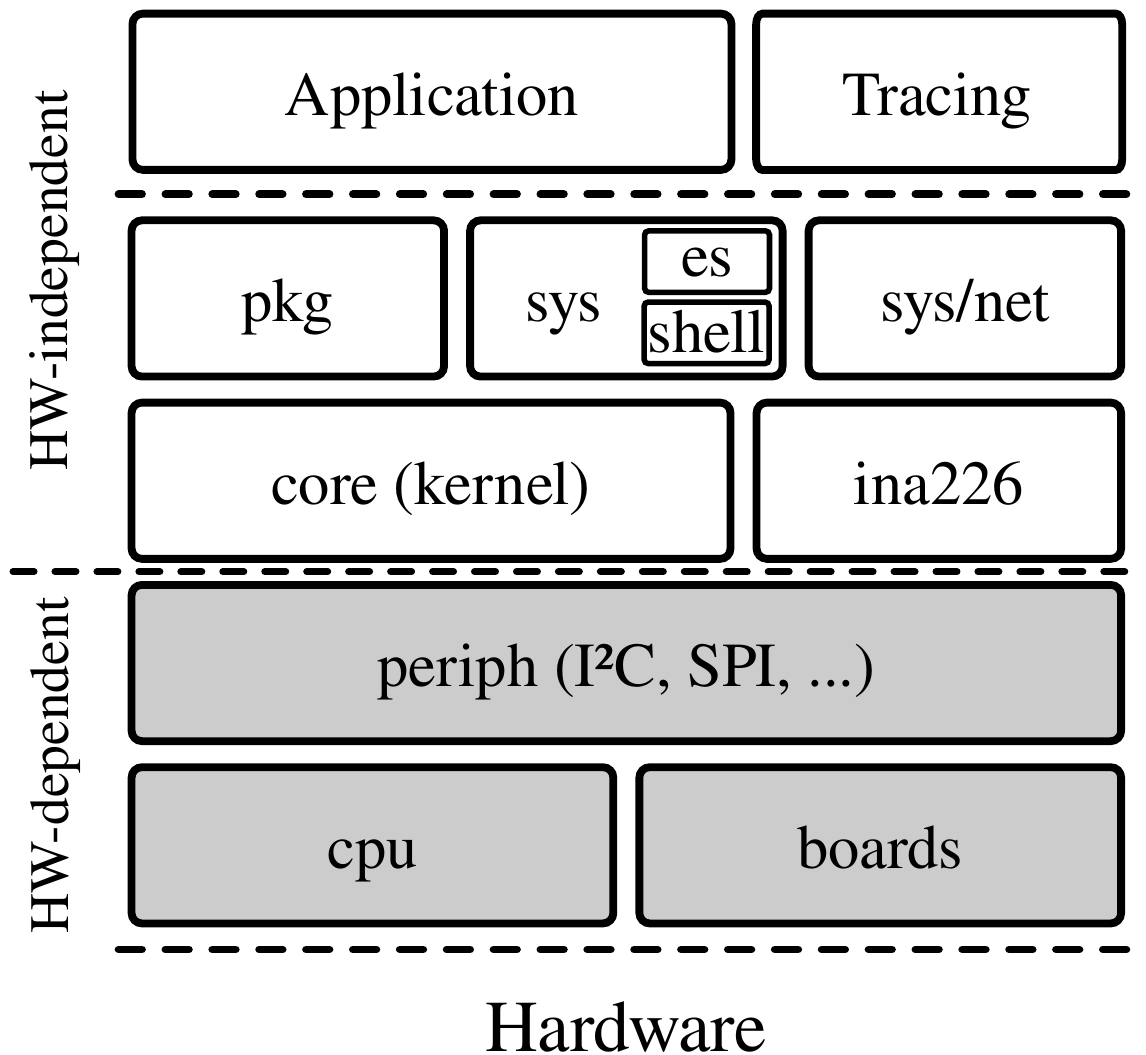}
	\caption{RIOT module architecture with integrated \texttt{es} command and application level tracing}
	\label{fig:riot_structure}
\end{figure}

In the second part, we discuss the RIOT software integration of energy measurement primitives.
RIOT is an open source operating system for constrained IoT devices focusing on open network standards and well-known programming interfaces.
Software sitting on top of the RIOT operating system can be moved to other platforms without requiring changes to the source code thanks to its extensive hardware abstraction.
It supports multi-threading with priority based tickless scheduling.
A major benefit of using RIOT comes with portability and extensive library and driver support.
Thereby it enables fast development of highly modular and loosely coupled application designs leveraging reusable code that is not tied to specific hardware.

Two orthogonal concepts are implemented for the online measurement that provide different trade-offs between integration complexity and accuracy.
To add energy awareness to new applications, we provide an API to explicitly gauge the energy consumed on an individual task base i.e. for predefined code-sections.
Additionally, we implement an extension to the OS scheduler to provides energy statistics on a per-thread level.
With this implicit consumption attribution to software entities, the system can even provide energy usage information without requiring any changes to the application code.

\subsection{Hardware}
The hardware is composed of an off-the-shelf evaluation board, a super capacitor as energy buffer and a measurement module to quantify the charging and discharging rate.
The entire setup is built as an orchestration of independent modules.
All of them can be exchanged, which leaves the design flexible regarding the selection of specific components.
To represent typical IoT use cases, an IO-interface serves connectivity to external peripherals like sensors for data acquisition.
A network uplink is provided by an IEEE 802.15.4 module based on the AT86RF233 radio chip.
When there is no network access available, a micro-SD card provides cost effective persistent storage for long term data logging.

A bare development board with the low-power STM32L476 MCU is used as base.
But the setup is not tied to this specific development board, as we will show later.
Using a bare development platform facilitates the use of different radio modules, voltage regulators, energy storage elements, or sensors, while keeping low level control.
Analyses of specific component properties is thereby simplified.

For the power measurement we design a configurable, yet simple module with the Texas Instruments INA226 shunt monitor.
It is interfaced via I$^2$C and allows changing the slave address to use more than one module in a single setup.
With a fixed maximum burden voltage of \SI{81.92}{\milli\volt} the measurement range can be manually switched based on three different shunt resistors for simple adaption.
Enabling a high dynamic range can be achieved by using an additional module with a higher value shunt resistor in combination with a bypass switch as shown in \cite{kzssk-ipcms-17}.

The assembly of all interconnected parts for the deployment-ready Energy-Harvesting system are depicted in \autoref{fig:eh_motherboard}.
The bare development board \circled{A} with the external IEEE 802.15.4 radio device \circled{B} on top is plugged into an interface board which in turn provides connection to persistent storage via a micro-SD slot \circled{C}, the power subsystem and various external sensors.
A \SI{100}{\farad} (\SI{2.7}{\volt}) super capacitor \circled{D} in combination with a custom photovoltaic charging circuit \circled{E} and the previously introduced measurement module \circled{F} form the external modular power subsystem.
All components like the radio, micro-SD card and external sensors can be powered down completely by individually switchable transistors. These are integrated on the interconnect-board below the MCU and are controlled by plain GPIO control. 

\subsection{Integrated System Software}

Integration into RIOT is achieved by providing a peripheral driver for controlling the INA226 over I$^2$C via the respective interfaces of the hardware abstraction layer.
Apart from raw register access the driver provides functions for conversion to physical units and calibration. 
Required calibration values are gathered by connecting a trusted reference multimeter and the measurement module to the same load of a test application. The calibration is enabled by then feeding the  values into compile time configuration.

We extend the simple command line interface of RIOT with an additional command named \texttt{es}.
It builds on top of the existing \texttt{ps} command and adds information on power draw and energy consumed per thread, similar to the default statistics like stack usage and context switching count.
The logic for that is implemented with a separate background thread that controls the measurement, reads samples from the external module, and performs required calculations.
By that, the thread priority control of the OS can be used to adjust between precise timeliness of the measurements and less invasiveness.

The attribution schema splits the samples according to the time each thread was active and accounts it to the different involved threads.
This approach has the benefit that not a single line of code needs to be changed to give an overview on energy expenditure by different parts of the application.
However, depending on application properties this method may be inaccurate.
Energy consumed by threads that trigger an action of high energy demand and then go back to sleep immediately, cannot be evaluated very accurately by this approach.

To overcome this limitation, a tracing mechanism is implemented to let an application explicitly record energy traces of specific tasks defined by the developer.
Tracing can be controlled by the interface calls \texttt{trace\_start()} and \texttt{trace\_stop()}. 
It may record a full series of energy samples or return a single aggregated value.
Using this tracing, an application can re-evaluate its energy use per task in varying conditions like ambient temperature, state of charge, or degraded component health.
When the measurement function is not in use, the module is put to a low-power mode and consumes less than \SI{2}{\micro\ampere}.

\autoref{fig:riot_structure} visualizes a simplified RIOT software stack that integrates the aforementioned components.
The dashed line in the middle separates the stack into hardware dependent parts on the bottom and hardware independent parts above.
The lower line illustrates the border between OS modules and the specific hardware it is running on, and the upper line splits user- or application-code from the OS.
In particular, the driver of the INA226 is situated on top of the hardware abstraction and is thus available on every platform that supports I$^2$C.
The \texttt{es} command is implemented as a \texttt{sys} component using the shell module.
Next to the application the explicit tracing mechanism resides as an independent module.

%% file: evaluation.tex
\section{Evalution of In Situ Measurement}
\label{sec:evaluation}
We validate \sname on four largely diverse MCU-platforms and evaluate measurement accuracy and introduced overhead.
In particular, the configuration of the physical communication layer and runtime settings for the shunt monitor are considered.
Relevant adjustments regarding I$^2$C communication are the clock speed and the selection of suitable pull-up resistance.
Increasing the clock speed reduces the time spent for communication but demands for lower resistances to actually achieve the faster switching times.
On the other hand, the lower resistance raises the power loss for the time in which the signal lines are held low.
Apart from the electrical properties, we analyze the impact of module parameters that are adjustable at runtime.
For this, the sampling rate related settings consisting of conversion time and averaging are examined in different combinations.
Conversion time here refers to the duration of a single internal sampling step that may then be additionally averaged before being processed as a sample on the MCU.

\paragraph{Setup}
A Keithley DMM7510 bench multimeter is employed as an accurate ground truth for reference measurements.
To ensure high accuracy, all measurements are performed after the \SI{90}{\minute} warm-up period of the multimeter and an auto calibration cycle is performed in advance.
Measurements are repeatedly executed using automated scripts leveraging the remote control features of the employed bench multimeter and power supply (Siglent SPD3303C).
Accordingly, the sensor node is controlled by a custom RIOT-shell interface that can start test runs and handles configuration of all runtime parameters.

\begin{figure*}[t]
	\centering
	\begin{minipage}[t]{\columnwidth}
		\input{figs/rel_err_dist.pgf}
		\caption{Relative measurement error at constant load with \SI{332}{\micro\second} bus/shunt conversion time and 16 avg. steps (IQR: 25th-75th percentile, whiskers: Q1-1.5*IQR and Q3+1.5*IQR)}
		\label{fig:static_load_err_dist}
	\end{minipage}
	\qquad
	\centering
	\begin{minipage}[t]{\columnwidth}
		\input{figs/dev_full_range.pgf}
		\caption{Absolute measurement deviation in 40 mA range at constant load (2.5th to 97.5th percentile)}
		\label{fig:static_load_err_full_range}
	\end{minipage}%
\end{figure*}

\subsection{Measurement Errors}
Errors of the measurement and digitization process itself are quantified by comparing results for measurement of the same variable metric between the device under test and the multimeter ground truth values.
The digital resolution is deduced from values given in the data sheet combined with the properties of the shunt resistor.
Factors that contribute to the sampling latency are described further down in \autoref{subsec:overhead}.
For this we focus on the overall time needed for reading measurements without considering interrupt latency separately.
The bare interrupt latency is platform specific but insignificant compared to the read duration.

\subsubsection{Accuracy}
Testing the accuracy of the measurement is done using a fixed resistor as static load.
The applied voltage is varied for different current values.
Measurements of the device under test and the reference are run simultaneously and repeated 1000 times.

\autoref{fig:static_load_err_dist} shows the distribution of the relative errors for current measurements ranging from \SI{200}{\micro\ampere} to \SI{2}{\milli\ampere}.
The successive increase towards the lower end of the scale is attributed to bigger relative impact of noise and the digitization resolution.
For loads of \SI{200}{\micro\ampere} the median error stays very close to \SI{1}{\percent}, which is well suited for self measurement.
Currents higher than \SI{1.8}{\milli\ampere} are much more stable around a median of \SI{0.5}{\permille}.
Extending this measurement up to the full range of \SI{40}{\milli\ampere} (for the selected \SI{2}{\ohm} shunt) confirmed that the pattern of negligible deviation stays true for higher currents, as can be seen from the absolute measurement deviation shown in \autoref{fig:static_load_err_full_range}.
The plot also highlights that a function for error compensation can be derived with a simple linear fit.
We refrain from implementing that for now because the absolute measurement error has no significant impact.

\begin{figure*}[t]
	\centering
	\input{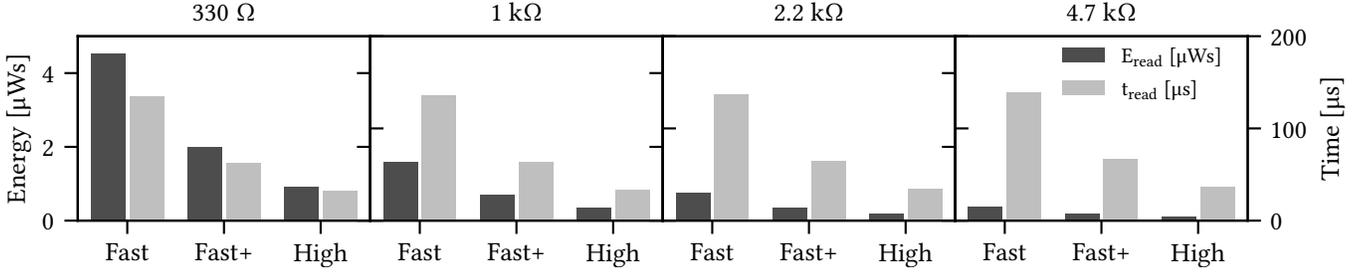}
	\caption{Energy and time overhead for a single register read at different I$^2$C configurations}
	\label{i2c_comm_power_time}
\end{figure*}

\subsubsection{Resolution}
The voltage resolution has a fixed value of \SI{1.25}{\milli\volt} by design.
Current measurement resolution depends on the active shunt resistor value ($R_{shunt}$) and the fixed \SI{2.5}{\micro\volt} LSB step size of the shunt voltage as indicated by \autoref{eq:i_lsb}.
\begin{equation} \label{eq:i_lsb}
I_{LSB} = \frac{\SI{2.5}{\micro\volt}}{R_{shunt}}
\end{equation}
We select a \SI{2}{\ohm} shunt that allows currents of over \SI{40}{\milli\ampere}, thus covering most typical IoT devices including connected sensors, while still maintaining a reasonable burden voltage around \SI{80}{\milli\volt}.
This shunt resistor value grants a resolution of \SI{1.25}{\micro\ampere}.
For tracing of the active node consumption this is considered accurate enough.
Though, for determination of low power sleep mode consumption additional provisioning of a separate measurement range would be beneficial.
In this work, we do not focus on this aspect as related work already covers this in great detail \cite{kzssk-ipcms-17, zx-nhfnp-13, sglll-mvehi-2017, jdcs-mpmem-07, dfpc-emfas-08}.

\subsubsection{Latency}
The dominating share in read latency is added by I$^2$C communication, which depends on the wiring and clock speed configuration.
Selecting an appropriate priority for the power measurement thread is also important.
Choosing the priority too low may starve the thread on high utilization and significantly increases latency and jitter.
With a high priority it is important to select the sampling rate of the module low enough to not be invasive to the application.
For deriving reasonable values, an application should be compiled for the target platform with the 'ps' command functionality included.
The overall computation overhead induced by the self-measurement can then be investigated and tweaked for the specific use-case and available hardware resources. 

\subsection{Cross-Platform Validation}
A major goal of \sname is to enable energy measurement on the majority of IoT platforms, which requires a seamless way to integrate the module with existing platforms.
We selected four largely heterogeneous boards to validate the cross-platform applicability of \sname.
For a representative set of samples from different manufacturers and architectures, we explicitly chose two devices from a higher performance class -- 32-Bit Cortex M4 based nucleo-l476rg and slstk3402a, one midrange device -- samr21-xpro running 32-Bit Cortex M0+, and the  8-Bit AVR8 arduino-mega2560 at the lower bound of the target devices.
With this choice, we spread our tests over three different manufacturers (STMicroelectronics, Silicon Labs, and Atmel/Microchip) and the two completely different architectures (ARM and AVR).
The names refer to the unique board names within RIOT.

With the chosen I$^2$C connectivity we already cover one hundred of the RIOT supported boards as of now and extending this list just requires to provide an I$^2$C peripheral driver.
Apart from the low-level driver, enabling \sname for a new platform only requires to physically connect six pins from the microcontroller to the measurement module, which in turn is connected to the energy storage.
Enabling the software support is done by providing pin mappings for the hardware (i.e. which pin of the MCU is connected to the measurement peripheral) and a dependency declaration for the driver module to the project Makefile.

While the general approach is not strictly tied to RIOT, some specific capabilities simplified the implementation.
The tick-less scheduling keeps the system load low in the absence of events.
With multi-threading the consumed energy can be implicitly attributed to separate parts of the application.
Also, prioritized scheduling provides direct control over the measurement invasiveness.

The general applicability of the proposed solution comes at the cost of a higher computational overhead compared to a highly specialized alternative but the following section reveals that the approach is still usable on low-end 8-bit platforms.
Considering that in the field of wireless sensor networks these architectures are even expected to be replaced with more capable 32-Bit variants \cite{kackz-sadpn-18}, we argue that the external module together with its tight OS integration can be considered generically applicable for the targeted devices.

\subsection{Overhead}
\label{subsec:overhead}
To quantify the overhead of the self measurement, we investigate the bus communication overhead in terms of time and power usage, power usage for the measurement itself and computation overhead to process the samples.
Power usage and computation complexity depend heavily on configuration parameters of the measurement module.
In contrast to that, the memory overhead depends mainly on the measurement application and is much bigger when a longer history of measurements or higher temporal resolution needs to be stored.
In this case an aggregated quantity like a moving average can save plenty of memory.

\subsubsection{Power}
The power overhead is generated by three contributors.
The measurement module consumes power during sampling, the bus spends energy on communication, and the shunt resistor introduces power loss.
Because I$^2$C is used, the related communication overhead directly depends on the configured bus clock and the wiring.
The maximum bus speed is only achievable with low value pull up resistors to ensure fast signal switching times.
Though, a lower resistance increases the power consumption when the line is pulled low, making it necessary to find an appropriate balance between high speed and low consumption.

The main factor to determine the necessary pull-up resistance is the bus capacitance ($C_b$) that needs to be charged by the current through the pull-up.
In the used setup $C_b$ is measured to be \SI{158}{\pico\farad}.
According to the I$^2$C bus specification, the maximum sufficient pull-up resistance can be calculated using \autoref{eq:rp_max}, where $t_r$ is the rise time specified by the respective I$^2$C speed.
\autoref{tbl:i2c_t_rise} lists the maximum resistor values for the given $C_b$ at the different bus speeds.

\begin{equation} \label{eq:rp_max}
R_p(max) = \frac{t_r}{0.8473 \cdot C_b}
\end{equation}

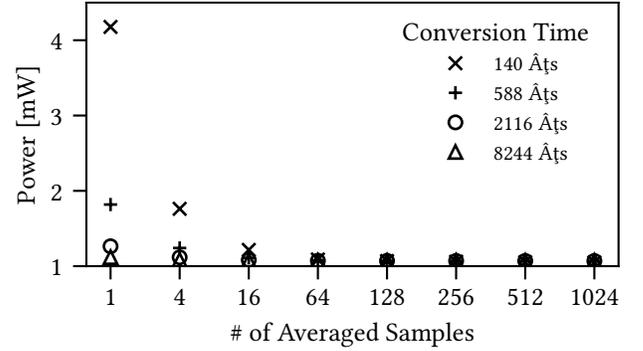
\begin{figure}
\centering
\input{figs/sample_power.pgf}
\caption{Power consumption of the measurement module for different configurations (I$^2$C speed high)}
\label{bulkmeasure_test_1}
\end{figure}

Accordingly, \autoref{i2c_comm_power_time} depicts the overhead caused by reading values from the external measurement module in the time and energy domain for multiple bus configurations.
Both, energy and time refer to a single register read.
The energy values contain the supply for the module itself as well as losses through pull-ups.
Plotted time values include all delays introduced by layers up to the reading application.
The changed parameters are the speed mode of the MCU I$^2$C peripheral and the pull-up configurations of the wiring.
It can be deduced from the data that selecting the pull-up values according to the specification consumes significantly more energy, while the time spent for communication doesn't decrease with a similar magnitude.
The results can be used to select the most energy efficient configuration from the overall system view.
Our measurements show that the energy cost per read can vary from \SI{4.5}{\micro\watt\second} (\SI{330}{\ohm} pull-up, fast mode) down to \SI{100}{\nano\watt\second} (\SI{4.7}{\kilo\ohm} pull-up, high mode).
Because the read duration does not improve with the same ratio, the energetic optimum is between those two points and also depends on how much energy is spent by the MCU during this transaction.
With \autoref{eq:e_complete} we identify the high speed configuration with a \SI{2.2}{\kilo\ohm} pull-up as most energy efficient for an average MCU consumption of \SI{12}{\milli\ampere}.
This stays in contrast to the I$^2$C specification recommendations and effectively leverages safety margins of the by imposing stricter wiring requirements.

\begin{equation} \label{eq:e_complete}
E_{read} = (P_{MCU} + P_{read}) \cdot t_{read}
\end{equation}

\begin{table}[]
	\centering
    \caption{Pull-up resistances based on I$^2$C specification}
	\vspace{0.25cm}
	\label{tbl:i2c_t_rise}
	\begin{tabular}{lrrll}
	\toprule
		I$^2$C mode & \multicolumn{1}{c}{$t_r$} & \multicolumn{1}{c}{$R_p(max)$}  \\
		\midrule
		Fast  & \SI{300}{\nano\second} & \SI{~2200}{\ohm} \\
		Fast+ & \SI{120}{\nano\second} & \SI{~900}{\ohm}  \\
		High  & \SI{40}{\nano\second}  & \SI{~300}{\ohm}  \\ 
		\bottomrule
	\end{tabular}
\end{table}

Using this configuration, we check whether different conversion times or averaging steps show a significant impact on the energy usage.
Respectively, \autoref{bulkmeasure_test_1} shows the average power consumption while sampling.

Short conversion times together with low averaging (\ie a high sampling frequency) shows significant impact compared to the overall measurement consumption.
With 16 averaging steps the overhead is only visible for the fastest sampling configuration which becomes completely negligible when 64 averaging steps or more are used.
Comparing the values with no averaging to the equivalent value with four averaging steps shows no noticeable dependence on how a targeted sampling interval is achieved.

The shunt resistor introduces a power loss that depends on the current drawn from the sensor node and the supply voltage.
A lower supply voltage implies the fixed maximum drop across the shunt has higher relative impact.
For a supply voltage of \SI{2.7}{\volt} and a load of \SI{10}{\milli\ampere} the power wasted in the shunt amounts to \SI{0.2}{\milli\watt} corresponding to around \SI{0.75}{\percent} of the overall power used.
At \SI{1}{\volt} using the maximum current of \SI{40}{\milli\ampere} the loss equals \SI{3.2}{\milli\watt} which already corresponds to \SI{8}{\percent}.
For systems using such low voltage either a smaller shunt can be used -- reducing the resolution -- or another measurement component with higher gain is~needed.
So for systems using voltages as low as \SI{1}{\volt} we recommend using a lower resistance shunt to trade efficiency for resolution.
Concluding the evaluation of the different power overhead contributors, yields the static consumption of the sampling process itself as the dominant factor for most cases.
The module communication becomes only relevant for very fast sampling and the shunt losses are only of concern when very low voltages and high resolution are required at the same time.
Compensation for the measurement consumption can be achieved by subtracting a fixed offset in case of medium to high sampling rates.
Maximum sampling rates require a linear correction taking the communication overhead into account.

\begin{table*}
	\begin{threeparttable}
		\centering
		\caption{Performance comparison of selected measurement solutions}
		\vspace{0.25cm}
		\label{tbl:comparison}
		\begin{tabular}{lccccr}
			\toprule
			& \multicolumn{1}{c}{Voltage} & \multicolumn{1}{c}{Accuracy} & \multicolumn{1}{c}{Power Overhead}  & \multicolumn{1}{c}{CPU Overhead} & \multicolumn{1}{c}{Read Duration}\\
			\midrule
			SPOT \cite{jdcs-mpmem-07}  & \xmark & \SI{3}{\percent} & $\sim$ \SI{5}{\milli\watt} 
			& \emph{unpublished}\tnote{*} & \emph{unpublished}\tnote{*}\\
			iCount \cite{dfpc-emfas-08} & \xmark & \SI{\pm15}{\percent} (\SI{5}{\micro\ampere} - \SI{50}{\milli\ampere}) & 
			\SI{0.3}{\micro\watt}  - \SI{30}{\micro\watt} 
			& \emph{unpublished}\tnote{\textdagger} & \SI{15}{\micro\second}\\
			Nemo \cite{zx-nhfnp-13} & \xmark & \SI{1.34}{\percent} avg. \SI{8}{\percent} max. & \SI{0.4}{\micro\watt} - \SI{12}{\milli\watt}
			& \SI{0.6}{\percent} (\SI{10}{\second}@\SI{0.5}{\hertz}, \SI{10}{\milli\second}@\SI{8}{\kilo\hertz}) & \SI{1}{\milli\second}\tnote{\textdaggerdbl}\\ 
			\sname & \cmark & < \SI{1}{\percent} (> \SI{0.2}{\milli\ampere}) & 
		    \SI{6}{\micro\watt} - \SI{1.1}{\milli\watt} 
			& \SI{0.16}{\permille}@\SI{1}{\hertz} \SI{1.5}{\percent}@\SI{100}{\hertz} & \SI{33}{\micro\second}\\
			\bottomrule
		\end{tabular}
		\begin{tablenotes}\footnotesize
			\item[*] A longer duration for reading samples is indicated because multiple steps are required per sample and the $I^2C$ is running with a slower clock (\SI{100}{\kilo\hertz})
			\item[\textdagger] By using an internal hardware counter the overhead is assumed to be very low, but hardware dependent
			\item[\textdaggerdbl] Best case for stated bandwidth, ignoring protocol overhead, assuming 16 bit samples
		\end{tablenotes}
	\end{threeparttable}
	\label{table2}
\end{table*}

\subsubsection{Computation}
\begin{figure}
	\centering
	\input{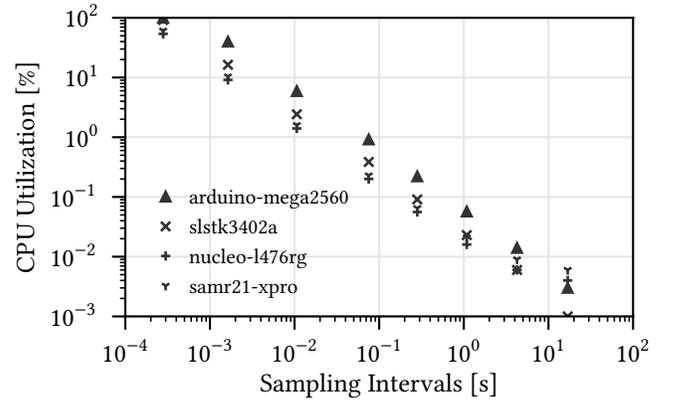}
	\caption{CPU utilization for different sampling intervals}
	\label{fig:cpu_util}
\end{figure}

We need to quantify the CPU occupancy while collecting the samples from the measurement module in order to select an appropriate sampling rate for the target domain.
When the CPU is busy with time sensitive computation, a high sampling rate and its overhead can be  invasive.
Conversely, for an application that only waits for an external sensor measurement to finish and is interested in a fine-grained power trace while the sensor is running, a high overhead for sampling is tolerable.
To measure the CPU utilization, the \texttt{ps} command of the RIOT shell is used.
The module is set to continuous sampling configuration with interrupt assertion enabled.
A dedicated thread for the measurement is woken up by each interrupt and reads the values for voltage and current.
This operating mode is left running for ten minutes before the \texttt{ps} command is executed to list the percentage of active time for all threads.
The values include every step from reading the device registers to physical unit conversion.
The whole procedure is run for eight different sample interval configurations.
\autoref{fig:cpu_util} shows the results of this evaluation on four exemplary platforms ranging from a 32-Bit Cortex-M4 board running at \SI{80}{\mega\hertz} (nucleo-l476rg) down to an 8-Bit AVR running at \SI{16}{\mega\hertz} (arduino-mega2560).
While the more capable nucleo-l476rg board stays at \SI{54}{\percent} utilization for a sampling time of \SI{280}{\micro\second}, the arduino-mega2560 is working on its limits and is not able to allocate any CPU resources to other application threads.
All values drop linearly over almost the full range down to \SI{0.16}{\permille} and \SI{0.57}{\permille} respectively at around \SI{1}{\second}.
It is noteworthy that exhausted computational power at the highest sampling rate does not prohibit using \sname on that platform.
Limited computational capabilities are also expected to decrease the required sampling rate due to the lower rate of events that need to be measured.

Comparing these values with the raw computational performance of the MCUs shows that the utilization on arduino-mega2560 only increases by a factor $\approx 4$ compared to the nucleo-l476rg whereas the performance differs by a factor of 6 (16 million instructions per second (MIPS) vs. 100 MIPS).
The CPU utilization is thus dominated by the I$^2$C transfer time and not limited by the raw instruction performance.
As the slowest sampling rate only generates 35 context switches to the measurement thread, its CPU values are subject to inaccuracies of the \texttt{ps} command.

Using the module's integrated power calculation instead of reading current and voltage separately can about halve the time for I$^2$C reading.
More room for improvement exists in the I$^2$C peripheral drivers of RIOT, which often do not leverage all the hardware features such as interrupt control or direct memory access.

\subsubsection{Memory}
Memory overhead can be considered from three different perspectives.
The static memory used by the module driver, the measurement thread, and a buffer for storing a history if necessary.
The driver itself uses up to \SI{560}{\byte}, depending on the actually used functionality.
The application uses additional \SI{36}{\byte} for the ISR callback, \SI{32}{\byte} for a message queue, \SI{268}{\byte} for measurement thread code and additional \SI{256}{\byte} of thread stack size, but the overall memory requirement slightly differs with target architecture and enabled features.
Depending on whether separate voltage and current samples need to be stored or a single power value is enough, a history needs either additional two or four bytes per sample.

\subsection{Discussion: Rating \sname}
\sname was designed as a highly portable approach that can be deployed with over one hundred different boards using a hardware shield carrying only commodity components, and software tightly integrated with the RIOT OS.
In contrast to many existing solutions, \sname neither exploits special aspects of individual boards, nor takes advantage of highly specialized or expensive components. Correspondingly, it is natural to question its qualitative positioning with respect to the current state of the art as outlined in \autoref{sec:problem}.

We compare the key performance indicators of \sname and those solutions that are roughly comparable and have published corresponding measurements in  \autoref{tbl:comparison}.
The indicators are \one~{availability of voltage readings}, \two~{relative errors}, \three~{overhead in power and CPU utilization}, and \four~{reading performance}.
Overall we find \sname in the vicinity of the best performers for each indicator.
In particular, our system appears to be the most balanced with respect to the considered metrics, and it is the only solution that actually measures voltage.

SPOT \cite{jdcs-mpmem-07} and iCount \cite{dfpc-emfas-08} do not keep track of voltage and instead assume a constant value.
In particular for super capacitor based energy-harvesting systems, this assumption is not valid and leads to erroneous results.
Accessing measurement values of SPOT is also done via I$^2$C, but the related communication overhead (in terms of power and CPU utilization) is not evaluated.
In contrast, \sname takes  I$^2$C cost carefully into consideration and can adapt to its different configurations.

By using an internal hardware counter and a signal of the switching regulator, iCount almost completely removes the need for additional hardware components and hence has low power overhead.
This comes at the cost of losing general applicability for different supply circuits and has the vital downside of requiring complex per device calibration.
It is also noteworthy that supply circuit properties and sampling rate affect the achievable resolution \cite{zx-nhfnp-13}.

Nemo \cite{zx-nhfnp-13} brings an additional MCU-board which guarantees good measurement accuracy, high sampling rates, and relatively low CPU overhead.
The major downsides are its high complexity and power overhead.
With \SI{150}{\micro\ampere} in sleep mode it already consumes almost halve the power of our setup during measurements.
In active operation this increases to \SI{4.6}{\milli\ampere}, effectively requiring an additional power supply for long term in situ deployments.

Nemo is focused on seamless integration into one specific platform and avoiding additional wiring by using modulation of the supply voltage to transmit its information to the measured node.
The authors give an example of the overhead based on a use case of sampling at \SI{0.5}{\hertz} for \SI{10}{\second} and then sampling at \SI{8}{\kilo\hertz} for \SI{10}{\milli\second} while buffering all data on the measurement node.
Transmitting the buffer to the observed node afterwards results in an average overhead of \SI{0.6}{\percent}.
The communication bandwidth implies that the achievable continuous meter-to-node data rate is over 60 times slower than with the $I^2C$ configuration we are using.
For use cases where the node continuously needs timely information of its power consumption (\eg when correlating power usage with software tasks), this slower data rate combined with buffering is inadequate.

From these observations, we conclude that specifically optimized custom solutions for specific use cases can slightly outperform our generic approach, but the commodity design of \sname proves to be significantly more versatile without sacrificing performance.

%% file: figs/dev_full_range.pgf
\begingroup%
\makeatletter%
\begin{pgfpicture}%
\pgfpathrectangle{\pgfpointorigin}{\pgfqpoint{3.347508in}{2.222224in}}%
\pgfusepath{use as bounding box, clip}%
\begin{pgfscope}%
\pgfsetbuttcap%
\pgfsetmiterjoin%
\definecolor{currentfill}{rgb}{1.000000,1.000000,1.000000}%
\pgfsetfillcolor{currentfill}%
\pgfsetlinewidth{0.000000pt}%
\definecolor{currentstroke}{rgb}{1.000000,1.000000,1.000000}%
\pgfsetstrokecolor{currentstroke}%
\pgfsetdash{}{0pt}%
\pgfpathmoveto{\pgfqpoint{0.000000in}{0.000000in}}%
\pgfpathlineto{\pgfqpoint{3.347508in}{0.000000in}}%
\pgfpathlineto{\pgfqpoint{3.347508in}{2.222224in}}%
\pgfpathlineto{\pgfqpoint{0.000000in}{2.222224in}}%
\pgfpathclose%
\pgfusepath{fill}%
\end{pgfscope}%
\begin{pgfscope}%
\pgfsetbuttcap%
\pgfsetmiterjoin%
\definecolor{currentfill}{rgb}{1.000000,1.000000,1.000000}%
\pgfsetfillcolor{currentfill}%
\pgfsetlinewidth{0.000000pt}%
\definecolor{currentstroke}{rgb}{0.000000,0.000000,0.000000}%
\pgfsetstrokecolor{currentstroke}%
\pgfsetstrokeopacity{0.000000}%
\pgfsetdash{}{0pt}%
\pgfpathmoveto{\pgfqpoint{0.423611in}{0.409611in}}%
\pgfpathlineto{\pgfqpoint{3.347508in}{0.409611in}}%
\pgfpathlineto{\pgfqpoint{3.347508in}{2.222224in}}%
\pgfpathlineto{\pgfqpoint{0.423611in}{2.222224in}}%
\pgfpathclose%
\pgfusepath{fill}%
\end{pgfscope}%
\begin{pgfscope}%
\pgfsetbuttcap%
\pgfsetroundjoin%
\definecolor{currentfill}{rgb}{0.000000,0.000000,0.000000}%
\pgfsetfillcolor{currentfill}%
\pgfsetfillopacity{0.000000}%
\pgfsetlinewidth{0.803000pt}%
\definecolor{currentstroke}{rgb}{0.000000,0.000000,0.000000}%
\pgfsetstrokecolor{currentstroke}%
\pgfsetdash{}{0pt}%
\pgfsys@defobject{currentmarker}{\pgfqpoint{0.000000in}{-0.069444in}}{\pgfqpoint{0.000000in}{0.000000in}}{%
\pgfpathmoveto{\pgfqpoint{0.000000in}{0.000000in}}%
\pgfpathlineto{\pgfqpoint{0.000000in}{-0.069444in}}%
\pgfusepath{stroke,fill}%
}%
\begin{pgfscope}%
\pgfsys@transformshift{0.543158in}{0.409611in}%
\pgfsys@useobject{currentmarker}{}%
\end{pgfscope}%
\end{pgfscope}%
\begin{pgfscope}%
\definecolor{textcolor}{rgb}{0.000000,0.000000,0.000000}%
\pgfsetstrokecolor{textcolor}%
\pgfsetfillcolor{textcolor}%
\pgftext[x=0.543158in,y=0.291555in,,top]{\color{textcolor}\rmfamily\fontsize{9.000000}{10.800000}\selectfont \(\displaystyle 0\)}%
\end{pgfscope}%
\begin{pgfscope}%
\pgfsetbuttcap%
\pgfsetroundjoin%
\definecolor{currentfill}{rgb}{0.000000,0.000000,0.000000}%
\pgfsetfillcolor{currentfill}%
\pgfsetfillopacity{0.000000}%
\pgfsetlinewidth{0.803000pt}%
\definecolor{currentstroke}{rgb}{0.000000,0.000000,0.000000}%
\pgfsetstrokecolor{currentstroke}%
\pgfsetdash{}{0pt}%
\pgfsys@defobject{currentmarker}{\pgfqpoint{0.000000in}{-0.069444in}}{\pgfqpoint{0.000000in}{0.000000in}}{%
\pgfpathmoveto{\pgfqpoint{0.000000in}{0.000000in}}%
\pgfpathlineto{\pgfqpoint{0.000000in}{-0.069444in}}%
\pgfusepath{stroke,fill}%
}%
\begin{pgfscope}%
\pgfsys@transformshift{1.211019in}{0.409611in}%
\pgfsys@useobject{currentmarker}{}%
\end{pgfscope}%
\end{pgfscope}%
\begin{pgfscope}%
\definecolor{textcolor}{rgb}{0.000000,0.000000,0.000000}%
\pgfsetstrokecolor{textcolor}%
\pgfsetfillcolor{textcolor}%
\pgftext[x=1.211019in,y=0.291555in,,top]{\color{textcolor}\rmfamily\fontsize{9.000000}{10.800000}\selectfont \(\displaystyle 10\)}%
\end{pgfscope}%
\begin{pgfscope}%
\pgfsetbuttcap%
\pgfsetroundjoin%
\definecolor{currentfill}{rgb}{0.000000,0.000000,0.000000}%
\pgfsetfillcolor{currentfill}%
\pgfsetfillopacity{0.000000}%
\pgfsetlinewidth{0.803000pt}%
\definecolor{currentstroke}{rgb}{0.000000,0.000000,0.000000}%
\pgfsetstrokecolor{currentstroke}%
\pgfsetdash{}{0pt}%
\pgfsys@defobject{currentmarker}{\pgfqpoint{0.000000in}{-0.069444in}}{\pgfqpoint{0.000000in}{0.000000in}}{%
\pgfpathmoveto{\pgfqpoint{0.000000in}{0.000000in}}%
\pgfpathlineto{\pgfqpoint{0.000000in}{-0.069444in}}%
\pgfusepath{stroke,fill}%
}%
\begin{pgfscope}%
\pgfsys@transformshift{1.878881in}{0.409611in}%
\pgfsys@useobject{currentmarker}{}%
\end{pgfscope}%
\end{pgfscope}%
\begin{pgfscope}%
\definecolor{textcolor}{rgb}{0.000000,0.000000,0.000000}%
\pgfsetstrokecolor{textcolor}%
\pgfsetfillcolor{textcolor}%
\pgftext[x=1.878881in,y=0.291555in,,top]{\color{textcolor}\rmfamily\fontsize{9.000000}{10.800000}\selectfont \(\displaystyle 20\)}%
\end{pgfscope}%
\begin{pgfscope}%
\pgfsetbuttcap%
\pgfsetroundjoin%
\definecolor{currentfill}{rgb}{0.000000,0.000000,0.000000}%
\pgfsetfillcolor{currentfill}%
\pgfsetfillopacity{0.000000}%
\pgfsetlinewidth{0.803000pt}%
\definecolor{currentstroke}{rgb}{0.000000,0.000000,0.000000}%
\pgfsetstrokecolor{currentstroke}%
\pgfsetdash{}{0pt}%
\pgfsys@defobject{currentmarker}{\pgfqpoint{0.000000in}{-0.069444in}}{\pgfqpoint{0.000000in}{0.000000in}}{%
\pgfpathmoveto{\pgfqpoint{0.000000in}{0.000000in}}%
\pgfpathlineto{\pgfqpoint{0.000000in}{-0.069444in}}%
\pgfusepath{stroke,fill}%
}%
\begin{pgfscope}%
\pgfsys@transformshift{2.546742in}{0.409611in}%
\pgfsys@useobject{currentmarker}{}%
\end{pgfscope}%
\end{pgfscope}%
\begin{pgfscope}%
\definecolor{textcolor}{rgb}{0.000000,0.000000,0.000000}%
\pgfsetstrokecolor{textcolor}%
\pgfsetfillcolor{textcolor}%
\pgftext[x=2.546742in,y=0.291555in,,top]{\color{textcolor}\rmfamily\fontsize{9.000000}{10.800000}\selectfont \(\displaystyle 30\)}%
\end{pgfscope}%
\begin{pgfscope}%
\pgfsetbuttcap%
\pgfsetroundjoin%
\definecolor{currentfill}{rgb}{0.000000,0.000000,0.000000}%
\pgfsetfillcolor{currentfill}%
\pgfsetfillopacity{0.000000}%
\pgfsetlinewidth{0.803000pt}%
\definecolor{currentstroke}{rgb}{0.000000,0.000000,0.000000}%
\pgfsetstrokecolor{currentstroke}%
\pgfsetdash{}{0pt}%
\pgfsys@defobject{currentmarker}{\pgfqpoint{0.000000in}{-0.069444in}}{\pgfqpoint{0.000000in}{0.000000in}}{%
\pgfpathmoveto{\pgfqpoint{0.000000in}{0.000000in}}%
\pgfpathlineto{\pgfqpoint{0.000000in}{-0.069444in}}%
\pgfusepath{stroke,fill}%
}%
\begin{pgfscope}%
\pgfsys@transformshift{3.214604in}{0.409611in}%
\pgfsys@useobject{currentmarker}{}%
\end{pgfscope}%
\end{pgfscope}%
\begin{pgfscope}%
\definecolor{textcolor}{rgb}{0.000000,0.000000,0.000000}%
\pgfsetstrokecolor{textcolor}%
\pgfsetfillcolor{textcolor}%
\pgftext[x=3.214604in,y=0.291555in,,top]{\color{textcolor}\rmfamily\fontsize{9.000000}{10.800000}\selectfont \(\displaystyle 40\)}%
\end{pgfscope}%
\begin{pgfscope}%
\definecolor{textcolor}{rgb}{0.000000,0.000000,0.000000}%
\pgfsetstrokecolor{textcolor}%
\pgfsetfillcolor{textcolor}%
\pgftext[x=1.885559in,y=0.125000in,,top]{\color{textcolor}\rmfamily\fontsize{9.000000}{10.800000}\selectfont Supply Current [mA]}%
\end{pgfscope}%
\begin{pgfscope}%
\pgfsetbuttcap%
\pgfsetroundjoin%
\definecolor{currentfill}{rgb}{0.000000,0.000000,0.000000}%
\pgfsetfillcolor{currentfill}%
\pgfsetfillopacity{0.000000}%
\pgfsetlinewidth{0.803000pt}%
\definecolor{currentstroke}{rgb}{0.000000,0.000000,0.000000}%
\pgfsetstrokecolor{currentstroke}%
\pgfsetdash{}{0pt}%
\pgfsys@defobject{currentmarker}{\pgfqpoint{-0.069444in}{0.000000in}}{\pgfqpoint{0.000000in}{0.000000in}}{%
\pgfpathmoveto{\pgfqpoint{0.000000in}{0.000000in}}%
\pgfpathlineto{\pgfqpoint{-0.069444in}{0.000000in}}%
\pgfusepath{stroke,fill}%
}%
\begin{pgfscope}%
\pgfsys@transformshift{0.423611in}{0.616058in}%
\pgfsys@useobject{currentmarker}{}%
\end{pgfscope}%
\end{pgfscope}%
\begin{pgfscope}%
\definecolor{textcolor}{rgb}{0.000000,0.000000,0.000000}%
\pgfsetstrokecolor{textcolor}%
\pgfsetfillcolor{textcolor}%
\pgftext[x=0.243055in,y=0.572683in,left,base]{\color{textcolor}\rmfamily\fontsize{9.000000}{10.800000}\selectfont \(\displaystyle 0\)}%
\end{pgfscope}%
\begin{pgfscope}%
\pgfsetbuttcap%
\pgfsetroundjoin%
\definecolor{currentfill}{rgb}{0.000000,0.000000,0.000000}%
\pgfsetfillcolor{currentfill}%
\pgfsetfillopacity{0.000000}%
\pgfsetlinewidth{0.803000pt}%
\definecolor{currentstroke}{rgb}{0.000000,0.000000,0.000000}%
\pgfsetstrokecolor{currentstroke}%
\pgfsetdash{}{0pt}%
\pgfsys@defobject{currentmarker}{\pgfqpoint{-0.069444in}{0.000000in}}{\pgfqpoint{0.000000in}{0.000000in}}{%
\pgfpathmoveto{\pgfqpoint{0.000000in}{0.000000in}}%
\pgfpathlineto{\pgfqpoint{-0.069444in}{0.000000in}}%
\pgfusepath{stroke,fill}%
}%
\begin{pgfscope}%
\pgfsys@transformshift{0.423611in}{0.955570in}%
\pgfsys@useobject{currentmarker}{}%
\end{pgfscope}%
\end{pgfscope}%
\begin{pgfscope}%
\definecolor{textcolor}{rgb}{0.000000,0.000000,0.000000}%
\pgfsetstrokecolor{textcolor}%
\pgfsetfillcolor{textcolor}%
\pgftext[x=0.180555in,y=0.912195in,left,base]{\color{textcolor}\rmfamily\fontsize{9.000000}{10.800000}\selectfont \(\displaystyle 10\)}%
\end{pgfscope}%
\begin{pgfscope}%
\pgfsetbuttcap%
\pgfsetroundjoin%
\definecolor{currentfill}{rgb}{0.000000,0.000000,0.000000}%
\pgfsetfillcolor{currentfill}%
\pgfsetfillopacity{0.000000}%
\pgfsetlinewidth{0.803000pt}%
\definecolor{currentstroke}{rgb}{0.000000,0.000000,0.000000}%
\pgfsetstrokecolor{currentstroke}%
\pgfsetdash{}{0pt}%
\pgfsys@defobject{currentmarker}{\pgfqpoint{-0.069444in}{0.000000in}}{\pgfqpoint{0.000000in}{0.000000in}}{%
\pgfpathmoveto{\pgfqpoint{0.000000in}{0.000000in}}%
\pgfpathlineto{\pgfqpoint{-0.069444in}{0.000000in}}%
\pgfusepath{stroke,fill}%
}%
\begin{pgfscope}%
\pgfsys@transformshift{0.423611in}{1.295083in}%
\pgfsys@useobject{currentmarker}{}%
\end{pgfscope}%
\end{pgfscope}%
\begin{pgfscope}%
\definecolor{textcolor}{rgb}{0.000000,0.000000,0.000000}%
\pgfsetstrokecolor{textcolor}%
\pgfsetfillcolor{textcolor}%
\pgftext[x=0.180555in,y=1.251708in,left,base]{\color{textcolor}\rmfamily\fontsize{9.000000}{10.800000}\selectfont \(\displaystyle 20\)}%
\end{pgfscope}%
\begin{pgfscope}%
\pgfsetbuttcap%
\pgfsetroundjoin%
\definecolor{currentfill}{rgb}{0.000000,0.000000,0.000000}%
\pgfsetfillcolor{currentfill}%
\pgfsetfillopacity{0.000000}%
\pgfsetlinewidth{0.803000pt}%
\definecolor{currentstroke}{rgb}{0.000000,0.000000,0.000000}%
\pgfsetstrokecolor{currentstroke}%
\pgfsetdash{}{0pt}%
\pgfsys@defobject{currentmarker}{\pgfqpoint{-0.069444in}{0.000000in}}{\pgfqpoint{0.000000in}{0.000000in}}{%
\pgfpathmoveto{\pgfqpoint{0.000000in}{0.000000in}}%
\pgfpathlineto{\pgfqpoint{-0.069444in}{0.000000in}}%
\pgfusepath{stroke,fill}%
}%
\begin{pgfscope}%
\pgfsys@transformshift{0.423611in}{1.634595in}%
\pgfsys@useobject{currentmarker}{}%
\end{pgfscope}%
\end{pgfscope}%
\begin{pgfscope}%
\definecolor{textcolor}{rgb}{0.000000,0.000000,0.000000}%
\pgfsetstrokecolor{textcolor}%
\pgfsetfillcolor{textcolor}%
\pgftext[x=0.180555in,y=1.591220in,left,base]{\color{textcolor}\rmfamily\fontsize{9.000000}{10.800000}\selectfont \(\displaystyle 30\)}%
\end{pgfscope}%
\begin{pgfscope}%
\pgfsetbuttcap%
\pgfsetroundjoin%
\definecolor{currentfill}{rgb}{0.000000,0.000000,0.000000}%
\pgfsetfillcolor{currentfill}%
\pgfsetfillopacity{0.000000}%
\pgfsetlinewidth{0.803000pt}%
\definecolor{currentstroke}{rgb}{0.000000,0.000000,0.000000}%
\pgfsetstrokecolor{currentstroke}%
\pgfsetdash{}{0pt}%
\pgfsys@defobject{currentmarker}{\pgfqpoint{-0.069444in}{0.000000in}}{\pgfqpoint{0.000000in}{0.000000in}}{%
\pgfpathmoveto{\pgfqpoint{0.000000in}{0.000000in}}%
\pgfpathlineto{\pgfqpoint{-0.069444in}{0.000000in}}%
\pgfusepath{stroke,fill}%
}%
\begin{pgfscope}%
\pgfsys@transformshift{0.423611in}{1.974108in}%
\pgfsys@useobject{currentmarker}{}%
\end{pgfscope}%
\end{pgfscope}%
\begin{pgfscope}%
\definecolor{textcolor}{rgb}{0.000000,0.000000,0.000000}%
\pgfsetstrokecolor{textcolor}%
\pgfsetfillcolor{textcolor}%
\pgftext[x=0.180555in,y=1.930733in,left,base]{\color{textcolor}\rmfamily\fontsize{9.000000}{10.800000}\selectfont \(\displaystyle 40\)}%
\end{pgfscope}%
\begin{pgfscope}%
\definecolor{textcolor}{rgb}{0.000000,0.000000,0.000000}%
\pgfsetstrokecolor{textcolor}%
\pgfsetfillcolor{textcolor}%
\pgftext[x=0.125000in,y=1.315917in,,bottom,rotate=90.000000]{\color{textcolor}\rmfamily\fontsize{9.000000}{10.800000}\selectfont Measurement Deviation [\(\displaystyle \upmu\)A]}%
\end{pgfscope}%
\begin{pgfscope}%
\pgfpathrectangle{\pgfqpoint{0.423611in}{0.409611in}}{\pgfqpoint{2.923897in}{1.812613in}}%
\pgfusepath{clip}%
\pgfsetbuttcap%
\pgfsetroundjoin%
\pgfsetlinewidth{0.752812pt}%
\definecolor{currentstroke}{rgb}{0.000000,0.000000,0.000000}%
\pgfsetstrokecolor{currentstroke}%
\pgfsetdash{}{0pt}%
\pgfpathmoveto{\pgfqpoint{0.569872in}{0.492002in}}%
\pgfpathlineto{\pgfqpoint{0.569872in}{0.590804in}}%
\pgfusepath{stroke}%
\end{pgfscope}%
\begin{pgfscope}%
\pgfpathrectangle{\pgfqpoint{0.423611in}{0.409611in}}{\pgfqpoint{2.923897in}{1.812613in}}%
\pgfusepath{clip}%
\pgfsetbuttcap%
\pgfsetroundjoin%
\pgfsetlinewidth{0.752812pt}%
\definecolor{currentstroke}{rgb}{0.000000,0.000000,0.000000}%
\pgfsetstrokecolor{currentstroke}%
\pgfsetdash{}{0pt}%
\pgfpathmoveto{\pgfqpoint{0.676730in}{0.573385in}}%
\pgfpathlineto{\pgfqpoint{0.676730in}{0.637278in}}%
\pgfusepath{stroke}%
\end{pgfscope}%
\begin{pgfscope}%
\pgfpathrectangle{\pgfqpoint{0.423611in}{0.409611in}}{\pgfqpoint{2.923897in}{1.812613in}}%
\pgfusepath{clip}%
\pgfsetbuttcap%
\pgfsetroundjoin%
\pgfsetlinewidth{0.752812pt}%
\definecolor{currentstroke}{rgb}{0.000000,0.000000,0.000000}%
\pgfsetstrokecolor{currentstroke}%
\pgfsetdash{}{0pt}%
\pgfpathmoveto{\pgfqpoint{0.810303in}{0.611337in}}%
\pgfpathlineto{\pgfqpoint{0.810303in}{0.675151in}}%
\pgfusepath{stroke}%
\end{pgfscope}%
\begin{pgfscope}%
\pgfpathrectangle{\pgfqpoint{0.423611in}{0.409611in}}{\pgfqpoint{2.923897in}{1.812613in}}%
\pgfusepath{clip}%
\pgfsetbuttcap%
\pgfsetroundjoin%
\pgfsetlinewidth{0.752812pt}%
\definecolor{currentstroke}{rgb}{0.000000,0.000000,0.000000}%
\pgfsetstrokecolor{currentstroke}%
\pgfsetdash{}{0pt}%
\pgfpathmoveto{\pgfqpoint{0.943875in}{0.681410in}}%
\pgfpathlineto{\pgfqpoint{0.943875in}{0.743725in}}%
\pgfusepath{stroke}%
\end{pgfscope}%
\begin{pgfscope}%
\pgfpathrectangle{\pgfqpoint{0.423611in}{0.409611in}}{\pgfqpoint{2.923897in}{1.812613in}}%
\pgfusepath{clip}%
\pgfsetbuttcap%
\pgfsetroundjoin%
\pgfsetlinewidth{0.752812pt}%
\definecolor{currentstroke}{rgb}{0.000000,0.000000,0.000000}%
\pgfsetstrokecolor{currentstroke}%
\pgfsetdash{}{0pt}%
\pgfpathmoveto{\pgfqpoint{1.077447in}{0.774197in}}%
\pgfpathlineto{\pgfqpoint{1.077447in}{0.847100in}}%
\pgfusepath{stroke}%
\end{pgfscope}%
\begin{pgfscope}%
\pgfpathrectangle{\pgfqpoint{0.423611in}{0.409611in}}{\pgfqpoint{2.923897in}{1.812613in}}%
\pgfusepath{clip}%
\pgfsetbuttcap%
\pgfsetroundjoin%
\pgfsetlinewidth{0.752812pt}%
\definecolor{currentstroke}{rgb}{0.000000,0.000000,0.000000}%
\pgfsetstrokecolor{currentstroke}%
\pgfsetdash{}{0pt}%
\pgfpathmoveto{\pgfqpoint{1.211019in}{0.787224in}}%
\pgfpathlineto{\pgfqpoint{1.211019in}{0.855049in}}%
\pgfusepath{stroke}%
\end{pgfscope}%
\begin{pgfscope}%
\pgfpathrectangle{\pgfqpoint{0.423611in}{0.409611in}}{\pgfqpoint{2.923897in}{1.812613in}}%
\pgfusepath{clip}%
\pgfsetbuttcap%
\pgfsetroundjoin%
\pgfsetlinewidth{0.752812pt}%
\definecolor{currentstroke}{rgb}{0.000000,0.000000,0.000000}%
\pgfsetstrokecolor{currentstroke}%
\pgfsetdash{}{0pt}%
\pgfpathmoveto{\pgfqpoint{1.344592in}{0.892081in}}%
\pgfpathlineto{\pgfqpoint{1.344592in}{0.954891in}}%
\pgfusepath{stroke}%
\end{pgfscope}%
\begin{pgfscope}%
\pgfpathrectangle{\pgfqpoint{0.423611in}{0.409611in}}{\pgfqpoint{2.923897in}{1.812613in}}%
\pgfusepath{clip}%
\pgfsetbuttcap%
\pgfsetroundjoin%
\pgfsetlinewidth{0.752812pt}%
\definecolor{currentstroke}{rgb}{0.000000,0.000000,0.000000}%
\pgfsetstrokecolor{currentstroke}%
\pgfsetdash{}{0pt}%
\pgfpathmoveto{\pgfqpoint{1.478164in}{0.972537in}}%
\pgfpathlineto{\pgfqpoint{1.478164in}{1.041823in}}%
\pgfusepath{stroke}%
\end{pgfscope}%
\begin{pgfscope}%
\pgfpathrectangle{\pgfqpoint{0.423611in}{0.409611in}}{\pgfqpoint{2.923897in}{1.812613in}}%
\pgfusepath{clip}%
\pgfsetbuttcap%
\pgfsetroundjoin%
\pgfsetlinewidth{0.752812pt}%
\definecolor{currentstroke}{rgb}{0.000000,0.000000,0.000000}%
\pgfsetstrokecolor{currentstroke}%
\pgfsetdash{}{0pt}%
\pgfpathmoveto{\pgfqpoint{1.611736in}{1.025501in}}%
\pgfpathlineto{\pgfqpoint{1.611736in}{1.128382in}}%
\pgfusepath{stroke}%
\end{pgfscope}%
\begin{pgfscope}%
\pgfpathrectangle{\pgfqpoint{0.423611in}{0.409611in}}{\pgfqpoint{2.923897in}{1.812613in}}%
\pgfusepath{clip}%
\pgfsetbuttcap%
\pgfsetroundjoin%
\pgfsetlinewidth{0.752812pt}%
\definecolor{currentstroke}{rgb}{0.000000,0.000000,0.000000}%
\pgfsetstrokecolor{currentstroke}%
\pgfsetdash{}{0pt}%
\pgfpathmoveto{\pgfqpoint{1.745309in}{1.136861in}}%
\pgfpathlineto{\pgfqpoint{1.745309in}{1.240421in}}%
\pgfusepath{stroke}%
\end{pgfscope}%
\begin{pgfscope}%
\pgfpathrectangle{\pgfqpoint{0.423611in}{0.409611in}}{\pgfqpoint{2.923897in}{1.812613in}}%
\pgfusepath{clip}%
\pgfsetbuttcap%
\pgfsetroundjoin%
\pgfsetlinewidth{0.752812pt}%
\definecolor{currentstroke}{rgb}{0.000000,0.000000,0.000000}%
\pgfsetstrokecolor{currentstroke}%
\pgfsetdash{}{0pt}%
\pgfpathmoveto{\pgfqpoint{1.878881in}{1.266224in}}%
\pgfpathlineto{\pgfqpoint{1.878881in}{1.368426in}}%
\pgfusepath{stroke}%
\end{pgfscope}%
\begin{pgfscope}%
\pgfpathrectangle{\pgfqpoint{0.423611in}{0.409611in}}{\pgfqpoint{2.923897in}{1.812613in}}%
\pgfusepath{clip}%
\pgfsetbuttcap%
\pgfsetroundjoin%
\pgfsetlinewidth{0.752812pt}%
\definecolor{currentstroke}{rgb}{0.000000,0.000000,0.000000}%
\pgfsetstrokecolor{currentstroke}%
\pgfsetdash{}{0pt}%
\pgfpathmoveto{\pgfqpoint{2.012453in}{1.308994in}}%
\pgfpathlineto{\pgfqpoint{2.012453in}{1.378611in}}%
\pgfusepath{stroke}%
\end{pgfscope}%
\begin{pgfscope}%
\pgfpathrectangle{\pgfqpoint{0.423611in}{0.409611in}}{\pgfqpoint{2.923897in}{1.812613in}}%
\pgfusepath{clip}%
\pgfsetbuttcap%
\pgfsetroundjoin%
\pgfsetlinewidth{0.752812pt}%
\definecolor{currentstroke}{rgb}{0.000000,0.000000,0.000000}%
\pgfsetstrokecolor{currentstroke}%
\pgfsetdash{}{0pt}%
\pgfpathmoveto{\pgfqpoint{2.146025in}{1.368757in}}%
\pgfpathlineto{\pgfqpoint{2.146025in}{1.435641in}}%
\pgfusepath{stroke}%
\end{pgfscope}%
\begin{pgfscope}%
\pgfpathrectangle{\pgfqpoint{0.423611in}{0.409611in}}{\pgfqpoint{2.923897in}{1.812613in}}%
\pgfusepath{clip}%
\pgfsetbuttcap%
\pgfsetroundjoin%
\pgfsetlinewidth{0.752812pt}%
\definecolor{currentstroke}{rgb}{0.000000,0.000000,0.000000}%
\pgfsetstrokecolor{currentstroke}%
\pgfsetdash{}{0pt}%
\pgfpathmoveto{\pgfqpoint{2.279598in}{1.464500in}}%
\pgfpathlineto{\pgfqpoint{2.279598in}{1.534439in}}%
\pgfusepath{stroke}%
\end{pgfscope}%
\begin{pgfscope}%
\pgfpathrectangle{\pgfqpoint{0.423611in}{0.409611in}}{\pgfqpoint{2.923897in}{1.812613in}}%
\pgfusepath{clip}%
\pgfsetbuttcap%
\pgfsetroundjoin%
\pgfsetlinewidth{0.752812pt}%
\definecolor{currentstroke}{rgb}{0.000000,0.000000,0.000000}%
\pgfsetstrokecolor{currentstroke}%
\pgfsetdash{}{0pt}%
\pgfpathmoveto{\pgfqpoint{2.413170in}{1.653269in}}%
\pgfpathlineto{\pgfqpoint{2.413170in}{1.759205in}}%
\pgfusepath{stroke}%
\end{pgfscope}%
\begin{pgfscope}%
\pgfpathrectangle{\pgfqpoint{0.423611in}{0.409611in}}{\pgfqpoint{2.923897in}{1.812613in}}%
\pgfusepath{clip}%
\pgfsetbuttcap%
\pgfsetroundjoin%
\pgfsetlinewidth{0.752812pt}%
\definecolor{currentstroke}{rgb}{0.000000,0.000000,0.000000}%
\pgfsetstrokecolor{currentstroke}%
\pgfsetdash{}{0pt}%
\pgfpathmoveto{\pgfqpoint{2.546742in}{1.752737in}}%
\pgfpathlineto{\pgfqpoint{2.546742in}{1.827803in}}%
\pgfusepath{stroke}%
\end{pgfscope}%
\begin{pgfscope}%
\pgfpathrectangle{\pgfqpoint{0.423611in}{0.409611in}}{\pgfqpoint{2.923897in}{1.812613in}}%
\pgfusepath{clip}%
\pgfsetbuttcap%
\pgfsetroundjoin%
\pgfsetlinewidth{0.752812pt}%
\definecolor{currentstroke}{rgb}{0.000000,0.000000,0.000000}%
\pgfsetstrokecolor{currentstroke}%
\pgfsetdash{}{0pt}%
\pgfpathmoveto{\pgfqpoint{2.680315in}{1.852562in}}%
\pgfpathlineto{\pgfqpoint{2.680315in}{1.921483in}}%
\pgfusepath{stroke}%
\end{pgfscope}%
\begin{pgfscope}%
\pgfpathrectangle{\pgfqpoint{0.423611in}{0.409611in}}{\pgfqpoint{2.923897in}{1.812613in}}%
\pgfusepath{clip}%
\pgfsetbuttcap%
\pgfsetroundjoin%
\pgfsetlinewidth{0.752812pt}%
\definecolor{currentstroke}{rgb}{0.000000,0.000000,0.000000}%
\pgfsetstrokecolor{currentstroke}%
\pgfsetdash{}{0pt}%
\pgfpathmoveto{\pgfqpoint{2.813887in}{2.007380in}}%
\pgfpathlineto{\pgfqpoint{2.813887in}{2.078678in}}%
\pgfusepath{stroke}%
\end{pgfscope}%
\begin{pgfscope}%
\pgfpathrectangle{\pgfqpoint{0.423611in}{0.409611in}}{\pgfqpoint{2.923897in}{1.812613in}}%
\pgfusepath{clip}%
\pgfsetbuttcap%
\pgfsetroundjoin%
\pgfsetlinewidth{0.752812pt}%
\definecolor{currentstroke}{rgb}{0.000000,0.000000,0.000000}%
\pgfsetstrokecolor{currentstroke}%
\pgfsetdash{}{0pt}%
\pgfpathmoveto{\pgfqpoint{2.947459in}{1.966978in}}%
\pgfpathlineto{\pgfqpoint{2.947459in}{2.039990in}}%
\pgfusepath{stroke}%
\end{pgfscope}%
\begin{pgfscope}%
\pgfpathrectangle{\pgfqpoint{0.423611in}{0.409611in}}{\pgfqpoint{2.923897in}{1.812613in}}%
\pgfusepath{clip}%
\pgfsetbuttcap%
\pgfsetroundjoin%
\pgfsetlinewidth{0.752812pt}%
\definecolor{currentstroke}{rgb}{0.000000,0.000000,0.000000}%
\pgfsetstrokecolor{currentstroke}%
\pgfsetdash{}{0pt}%
\pgfpathmoveto{\pgfqpoint{3.081031in}{1.918428in}}%
\pgfpathlineto{\pgfqpoint{3.081031in}{2.024016in}}%
\pgfusepath{stroke}%
\end{pgfscope}%
\begin{pgfscope}%
\pgfpathrectangle{\pgfqpoint{0.423611in}{0.409611in}}{\pgfqpoint{2.923897in}{1.812613in}}%
\pgfusepath{clip}%
\pgfsetbuttcap%
\pgfsetroundjoin%
\pgfsetlinewidth{0.752812pt}%
\definecolor{currentstroke}{rgb}{0.000000,0.000000,0.000000}%
\pgfsetstrokecolor{currentstroke}%
\pgfsetdash{}{0pt}%
\pgfpathmoveto{\pgfqpoint{3.214604in}{2.034524in}}%
\pgfpathlineto{\pgfqpoint{3.214604in}{2.139832in}}%
\pgfusepath{stroke}%
\end{pgfscope}%
\begin{pgfscope}%
\pgfpathrectangle{\pgfqpoint{0.423611in}{0.409611in}}{\pgfqpoint{2.923897in}{1.812613in}}%
\pgfusepath{clip}%
\pgfsetbuttcap%
\pgfsetroundjoin%
\definecolor{currentfill}{rgb}{0.000000,0.000000,0.000000}%
\pgfsetfillcolor{currentfill}%
\pgfsetfillopacity{0.000000}%
\pgfsetlinewidth{0.752812pt}%
\definecolor{currentstroke}{rgb}{0.000000,0.000000,0.000000}%
\pgfsetstrokecolor{currentstroke}%
\pgfsetdash{}{0pt}%
\pgfsys@defobject{currentmarker}{\pgfqpoint{-0.013889in}{-0.000000in}}{\pgfqpoint{0.013889in}{0.000000in}}{%
\pgfpathmoveto{\pgfqpoint{0.013889in}{-0.000000in}}%
\pgfpathlineto{\pgfqpoint{-0.013889in}{0.000000in}}%
\pgfusepath{stroke,fill}%
}%
\begin{pgfscope}%
\pgfsys@transformshift{0.569872in}{0.492002in}%
\pgfsys@useobject{currentmarker}{}%
\end{pgfscope}%
\begin{pgfscope}%
\pgfsys@transformshift{0.676730in}{0.573385in}%
\pgfsys@useobject{currentmarker}{}%
\end{pgfscope}%
\begin{pgfscope}%
\pgfsys@transformshift{0.810303in}{0.611337in}%
\pgfsys@useobject{currentmarker}{}%
\end{pgfscope}%
\begin{pgfscope}%
\pgfsys@transformshift{0.943875in}{0.681410in}%
\pgfsys@useobject{currentmarker}{}%
\end{pgfscope}%
\begin{pgfscope}%
\pgfsys@transformshift{1.077447in}{0.774197in}%
\pgfsys@useobject{currentmarker}{}%
\end{pgfscope}%
\begin{pgfscope}%
\pgfsys@transformshift{1.211019in}{0.787224in}%
\pgfsys@useobject{currentmarker}{}%
\end{pgfscope}%
\begin{pgfscope}%
\pgfsys@transformshift{1.344592in}{0.892081in}%
\pgfsys@useobject{currentmarker}{}%
\end{pgfscope}%
\begin{pgfscope}%
\pgfsys@transformshift{1.478164in}{0.972537in}%
\pgfsys@useobject{currentmarker}{}%
\end{pgfscope}%
\begin{pgfscope}%
\pgfsys@transformshift{1.611736in}{1.025501in}%
\pgfsys@useobject{currentmarker}{}%
\end{pgfscope}%
\begin{pgfscope}%
\pgfsys@transformshift{1.745309in}{1.136861in}%
\pgfsys@useobject{currentmarker}{}%
\end{pgfscope}%
\begin{pgfscope}%
\pgfsys@transformshift{1.878881in}{1.266224in}%
\pgfsys@useobject{currentmarker}{}%
\end{pgfscope}%
\begin{pgfscope}%
\pgfsys@transformshift{2.012453in}{1.308994in}%
\pgfsys@useobject{currentmarker}{}%
\end{pgfscope}%
\begin{pgfscope}%
\pgfsys@transformshift{2.146025in}{1.368757in}%
\pgfsys@useobject{currentmarker}{}%
\end{pgfscope}%
\begin{pgfscope}%
\pgfsys@transformshift{2.279598in}{1.464500in}%
\pgfsys@useobject{currentmarker}{}%
\end{pgfscope}%
\begin{pgfscope}%
\pgfsys@transformshift{2.413170in}{1.653269in}%
\pgfsys@useobject{currentmarker}{}%
\end{pgfscope}%
\begin{pgfscope}%
\pgfsys@transformshift{2.546742in}{1.752737in}%
\pgfsys@useobject{currentmarker}{}%
\end{pgfscope}%
\begin{pgfscope}%
\pgfsys@transformshift{2.680315in}{1.852562in}%
\pgfsys@useobject{currentmarker}{}%
\end{pgfscope}%
\begin{pgfscope}%
\pgfsys@transformshift{2.813887in}{2.007380in}%
\pgfsys@useobject{currentmarker}{}%
\end{pgfscope}%
\begin{pgfscope}%
\pgfsys@transformshift{2.947459in}{1.966978in}%
\pgfsys@useobject{currentmarker}{}%
\end{pgfscope}%
\begin{pgfscope}%
\pgfsys@transformshift{3.081031in}{1.918428in}%
\pgfsys@useobject{currentmarker}{}%
\end{pgfscope}%
\begin{pgfscope}%
\pgfsys@transformshift{3.214604in}{2.034524in}%
\pgfsys@useobject{currentmarker}{}%
\end{pgfscope}%
\end{pgfscope}%
\begin{pgfscope}%
\pgfpathrectangle{\pgfqpoint{0.423611in}{0.409611in}}{\pgfqpoint{2.923897in}{1.812613in}}%
\pgfusepath{clip}%
\pgfsetbuttcap%
\pgfsetroundjoin%
\definecolor{currentfill}{rgb}{0.000000,0.000000,0.000000}%
\pgfsetfillcolor{currentfill}%
\pgfsetfillopacity{0.000000}%
\pgfsetlinewidth{0.752812pt}%
\definecolor{currentstroke}{rgb}{0.000000,0.000000,0.000000}%
\pgfsetstrokecolor{currentstroke}%
\pgfsetdash{}{0pt}%
\pgfsys@defobject{currentmarker}{\pgfqpoint{-0.013889in}{-0.000000in}}{\pgfqpoint{0.013889in}{0.000000in}}{%
\pgfpathmoveto{\pgfqpoint{0.013889in}{-0.000000in}}%
\pgfpathlineto{\pgfqpoint{-0.013889in}{0.000000in}}%
\pgfusepath{stroke,fill}%
}%
\begin{pgfscope}%
\pgfsys@transformshift{0.569872in}{0.590804in}%
\pgfsys@useobject{currentmarker}{}%
\end{pgfscope}%
\begin{pgfscope}%
\pgfsys@transformshift{0.676730in}{0.637278in}%
\pgfsys@useobject{currentmarker}{}%
\end{pgfscope}%
\begin{pgfscope}%
\pgfsys@transformshift{0.810303in}{0.675151in}%
\pgfsys@useobject{currentmarker}{}%
\end{pgfscope}%
\begin{pgfscope}%
\pgfsys@transformshift{0.943875in}{0.743725in}%
\pgfsys@useobject{currentmarker}{}%
\end{pgfscope}%
\begin{pgfscope}%
\pgfsys@transformshift{1.077447in}{0.847100in}%
\pgfsys@useobject{currentmarker}{}%
\end{pgfscope}%
\begin{pgfscope}%
\pgfsys@transformshift{1.211019in}{0.855049in}%
\pgfsys@useobject{currentmarker}{}%
\end{pgfscope}%
\begin{pgfscope}%
\pgfsys@transformshift{1.344592in}{0.954891in}%
\pgfsys@useobject{currentmarker}{}%
\end{pgfscope}%
\begin{pgfscope}%
\pgfsys@transformshift{1.478164in}{1.041823in}%
\pgfsys@useobject{currentmarker}{}%
\end{pgfscope}%
\begin{pgfscope}%
\pgfsys@transformshift{1.611736in}{1.128382in}%
\pgfsys@useobject{currentmarker}{}%
\end{pgfscope}%
\begin{pgfscope}%
\pgfsys@transformshift{1.745309in}{1.240421in}%
\pgfsys@useobject{currentmarker}{}%
\end{pgfscope}%
\begin{pgfscope}%
\pgfsys@transformshift{1.878881in}{1.368426in}%
\pgfsys@useobject{currentmarker}{}%
\end{pgfscope}%
\begin{pgfscope}%
\pgfsys@transformshift{2.012453in}{1.378611in}%
\pgfsys@useobject{currentmarker}{}%
\end{pgfscope}%
\begin{pgfscope}%
\pgfsys@transformshift{2.146025in}{1.435641in}%
\pgfsys@useobject{currentmarker}{}%
\end{pgfscope}%
\begin{pgfscope}%
\pgfsys@transformshift{2.279598in}{1.534439in}%
\pgfsys@useobject{currentmarker}{}%
\end{pgfscope}%
\begin{pgfscope}%
\pgfsys@transformshift{2.413170in}{1.759205in}%
\pgfsys@useobject{currentmarker}{}%
\end{pgfscope}%
\begin{pgfscope}%
\pgfsys@transformshift{2.546742in}{1.827803in}%
\pgfsys@useobject{currentmarker}{}%
\end{pgfscope}%
\begin{pgfscope}%
\pgfsys@transformshift{2.680315in}{1.921483in}%
\pgfsys@useobject{currentmarker}{}%
\end{pgfscope}%
\begin{pgfscope}%
\pgfsys@transformshift{2.813887in}{2.078678in}%
\pgfsys@useobject{currentmarker}{}%
\end{pgfscope}%
\begin{pgfscope}%
\pgfsys@transformshift{2.947459in}{2.039990in}%
\pgfsys@useobject{currentmarker}{}%
\end{pgfscope}%
\begin{pgfscope}%
\pgfsys@transformshift{3.081031in}{2.024016in}%
\pgfsys@useobject{currentmarker}{}%
\end{pgfscope}%
\begin{pgfscope}%
\pgfsys@transformshift{3.214604in}{2.139832in}%
\pgfsys@useobject{currentmarker}{}%
\end{pgfscope}%
\end{pgfscope}%
\begin{pgfscope}%
\pgfpathrectangle{\pgfqpoint{0.423611in}{0.409611in}}{\pgfqpoint{2.923897in}{1.812613in}}%
\pgfusepath{clip}%
\pgfsetbuttcap%
\pgfsetroundjoin%
\pgfsetlinewidth{1.003750pt}%
\definecolor{currentstroke}{rgb}{0.500000,0.500000,0.500000}%
\pgfsetstrokecolor{currentstroke}%
\pgfsetdash{{3.700000pt}{1.600000pt}}{0.000000pt}%
\pgfpathmoveto{\pgfqpoint{0.556515in}{0.505353in}}%
\pgfpathlineto{\pgfqpoint{0.569872in}{0.513515in}}%
\pgfpathlineto{\pgfqpoint{0.583230in}{0.521677in}}%
\pgfpathlineto{\pgfqpoint{0.596587in}{0.529839in}}%
\pgfpathlineto{\pgfqpoint{0.609944in}{0.538001in}}%
\pgfpathlineto{\pgfqpoint{0.623301in}{0.546163in}}%
\pgfpathlineto{\pgfqpoint{0.636659in}{0.554325in}}%
\pgfpathlineto{\pgfqpoint{0.650016in}{0.562487in}}%
\pgfpathlineto{\pgfqpoint{0.663373in}{0.570649in}}%
\pgfpathlineto{\pgfqpoint{0.676730in}{0.578812in}}%
\pgfpathlineto{\pgfqpoint{0.810303in}{0.660432in}}%
\pgfpathlineto{\pgfqpoint{0.943875in}{0.742053in}}%
\pgfpathlineto{\pgfqpoint{1.077447in}{0.823674in}}%
\pgfpathlineto{\pgfqpoint{1.211019in}{0.905294in}}%
\pgfpathlineto{\pgfqpoint{1.344592in}{0.986915in}}%
\pgfpathlineto{\pgfqpoint{1.478164in}{1.068536in}}%
\pgfpathlineto{\pgfqpoint{1.611736in}{1.150156in}}%
\pgfpathlineto{\pgfqpoint{1.745309in}{1.231777in}}%
\pgfpathlineto{\pgfqpoint{1.878881in}{1.313398in}}%
\pgfpathlineto{\pgfqpoint{2.012453in}{1.395018in}}%
\pgfpathlineto{\pgfqpoint{2.146025in}{1.476639in}}%
\pgfpathlineto{\pgfqpoint{2.279598in}{1.558260in}}%
\pgfpathlineto{\pgfqpoint{2.413170in}{1.639881in}}%
\pgfpathlineto{\pgfqpoint{2.546742in}{1.721501in}}%
\pgfpathlineto{\pgfqpoint{2.680315in}{1.803122in}}%
\pgfpathlineto{\pgfqpoint{2.813887in}{1.884743in}}%
\pgfpathlineto{\pgfqpoint{2.947459in}{1.966363in}}%
\pgfpathlineto{\pgfqpoint{3.081031in}{2.047984in}}%
\pgfpathlineto{\pgfqpoint{3.214604in}{2.129605in}}%
\pgfusepath{stroke}%
\end{pgfscope}%
\begin{pgfscope}%
\pgfpathrectangle{\pgfqpoint{0.423611in}{0.409611in}}{\pgfqpoint{2.923897in}{1.812613in}}%
\pgfusepath{clip}%
\pgfsetbuttcap%
\pgfsetroundjoin%
\definecolor{currentfill}{rgb}{0.000000,0.000000,0.000000}%
\pgfsetfillcolor{currentfill}%
\pgfsetlinewidth{1.003750pt}%
\definecolor{currentstroke}{rgb}{0.000000,0.000000,0.000000}%
\pgfsetstrokecolor{currentstroke}%
\pgfsetdash{}{0pt}%
\pgfsys@defobject{currentmarker}{\pgfqpoint{-0.010417in}{-0.010417in}}{\pgfqpoint{0.010417in}{0.010417in}}{%
\pgfpathmoveto{\pgfqpoint{0.000000in}{-0.010417in}}%
\pgfpathcurveto{\pgfqpoint{0.002763in}{-0.010417in}}{\pgfqpoint{0.005412in}{-0.009319in}}{\pgfqpoint{0.007366in}{-0.007366in}}%
\pgfpathcurveto{\pgfqpoint{0.009319in}{-0.005412in}}{\pgfqpoint{0.010417in}{-0.002763in}}{\pgfqpoint{0.010417in}{0.000000in}}%
\pgfpathcurveto{\pgfqpoint{0.010417in}{0.002763in}}{\pgfqpoint{0.009319in}{0.005412in}}{\pgfqpoint{0.007366in}{0.007366in}}%
\pgfpathcurveto{\pgfqpoint{0.005412in}{0.009319in}}{\pgfqpoint{0.002763in}{0.010417in}}{\pgfqpoint{0.000000in}{0.010417in}}%
\pgfpathcurveto{\pgfqpoint{-0.002763in}{0.010417in}}{\pgfqpoint{-0.005412in}{0.009319in}}{\pgfqpoint{-0.007366in}{0.007366in}}%
\pgfpathcurveto{\pgfqpoint{-0.009319in}{0.005412in}}{\pgfqpoint{-0.010417in}{0.002763in}}{\pgfqpoint{-0.010417in}{0.000000in}}%
\pgfpathcurveto{\pgfqpoint{-0.010417in}{-0.002763in}}{\pgfqpoint{-0.009319in}{-0.005412in}}{\pgfqpoint{-0.007366in}{-0.007366in}}%
\pgfpathcurveto{\pgfqpoint{-0.005412in}{-0.009319in}}{\pgfqpoint{-0.002763in}{-0.010417in}}{\pgfqpoint{0.000000in}{-0.010417in}}%
\pgfpathclose%
\pgfusepath{stroke,fill}%
}%
\begin{pgfscope}%
\pgfsys@transformshift{0.569872in}{0.554424in}%
\pgfsys@useobject{currentmarker}{}%
\end{pgfscope}%
\begin{pgfscope}%
\pgfsys@transformshift{0.676730in}{0.604274in}%
\pgfsys@useobject{currentmarker}{}%
\end{pgfscope}%
\begin{pgfscope}%
\pgfsys@transformshift{0.810303in}{0.641936in}%
\pgfsys@useobject{currentmarker}{}%
\end{pgfscope}%
\begin{pgfscope}%
\pgfsys@transformshift{0.943875in}{0.711229in}%
\pgfsys@useobject{currentmarker}{}%
\end{pgfscope}%
\begin{pgfscope}%
\pgfsys@transformshift{1.077447in}{0.808678in}%
\pgfsys@useobject{currentmarker}{}%
\end{pgfscope}%
\begin{pgfscope}%
\pgfsys@transformshift{1.211019in}{0.819569in}%
\pgfsys@useobject{currentmarker}{}%
\end{pgfscope}%
\begin{pgfscope}%
\pgfsys@transformshift{1.344592in}{0.924515in}%
\pgfsys@useobject{currentmarker}{}%
\end{pgfscope}%
\begin{pgfscope}%
\pgfsys@transformshift{1.478164in}{1.007371in}%
\pgfsys@useobject{currentmarker}{}%
\end{pgfscope}%
\begin{pgfscope}%
\pgfsys@transformshift{1.611736in}{1.062731in}%
\pgfsys@useobject{currentmarker}{}%
\end{pgfscope}%
\begin{pgfscope}%
\pgfsys@transformshift{1.745309in}{1.179783in}%
\pgfsys@useobject{currentmarker}{}%
\end{pgfscope}%
\begin{pgfscope}%
\pgfsys@transformshift{1.878881in}{1.308912in}%
\pgfsys@useobject{currentmarker}{}%
\end{pgfscope}%
\begin{pgfscope}%
\pgfsys@transformshift{2.012453in}{1.341852in}%
\pgfsys@useobject{currentmarker}{}%
\end{pgfscope}%
\begin{pgfscope}%
\pgfsys@transformshift{2.146025in}{1.403680in}%
\pgfsys@useobject{currentmarker}{}%
\end{pgfscope}%
\begin{pgfscope}%
\pgfsys@transformshift{2.279598in}{1.501614in}%
\pgfsys@useobject{currentmarker}{}%
\end{pgfscope}%
\begin{pgfscope}%
\pgfsys@transformshift{2.413170in}{1.719116in}%
\pgfsys@useobject{currentmarker}{}%
\end{pgfscope}%
\begin{pgfscope}%
\pgfsys@transformshift{2.546742in}{1.787525in}%
\pgfsys@useobject{currentmarker}{}%
\end{pgfscope}%
\begin{pgfscope}%
\pgfsys@transformshift{2.680315in}{1.887049in}%
\pgfsys@useobject{currentmarker}{}%
\end{pgfscope}%
\begin{pgfscope}%
\pgfsys@transformshift{2.813887in}{2.040319in}%
\pgfsys@useobject{currentmarker}{}%
\end{pgfscope}%
\begin{pgfscope}%
\pgfsys@transformshift{2.947459in}{2.002739in}%
\pgfsys@useobject{currentmarker}{}%
\end{pgfscope}%
\begin{pgfscope}%
\pgfsys@transformshift{3.081031in}{1.979140in}%
\pgfsys@useobject{currentmarker}{}%
\end{pgfscope}%
\begin{pgfscope}%
\pgfsys@transformshift{3.214604in}{2.092781in}%
\pgfsys@useobject{currentmarker}{}%
\end{pgfscope}%
\end{pgfscope}%
\begin{pgfscope}%
\pgfsetrectcap%
\pgfsetmiterjoin%
\pgfsetlinewidth{0.803000pt}%
\definecolor{currentstroke}{rgb}{0.000000,0.000000,0.000000}%
\pgfsetstrokecolor{currentstroke}%
\pgfsetdash{}{0pt}%
\pgfpathmoveto{\pgfqpoint{0.423611in}{0.409611in}}%
\pgfpathlineto{\pgfqpoint{0.423611in}{2.222224in}}%
\pgfusepath{stroke}%
\end{pgfscope}%
\begin{pgfscope}%
\pgfsetrectcap%
\pgfsetmiterjoin%
\pgfsetlinewidth{0.803000pt}%
\definecolor{currentstroke}{rgb}{0.000000,0.000000,0.000000}%
\pgfsetstrokecolor{currentstroke}%
\pgfsetdash{}{0pt}%
\pgfpathmoveto{\pgfqpoint{3.347508in}{0.409611in}}%
\pgfpathlineto{\pgfqpoint{3.347508in}{2.222224in}}%
\pgfusepath{stroke}%
\end{pgfscope}%
\begin{pgfscope}%
\pgfsetrectcap%
\pgfsetmiterjoin%
\pgfsetlinewidth{0.803000pt}%
\definecolor{currentstroke}{rgb}{0.000000,0.000000,0.000000}%
\pgfsetstrokecolor{currentstroke}%
\pgfsetdash{}{0pt}%
\pgfpathmoveto{\pgfqpoint{0.423611in}{0.409611in}}%
\pgfpathlineto{\pgfqpoint{3.347508in}{0.409611in}}%
\pgfusepath{stroke}%
\end{pgfscope}%
\begin{pgfscope}%
\pgfsetrectcap%
\pgfsetmiterjoin%
\pgfsetlinewidth{0.803000pt}%
\definecolor{currentstroke}{rgb}{0.000000,0.000000,0.000000}%
\pgfsetstrokecolor{currentstroke}%
\pgfsetdash{}{0pt}%
\pgfpathmoveto{\pgfqpoint{0.423611in}{2.222224in}}%
\pgfpathlineto{\pgfqpoint{3.347508in}{2.222224in}}%
\pgfusepath{stroke}%
\end{pgfscope}%
\begin{pgfscope}%
\pgfsetbuttcap%
\pgfsetroundjoin%
\pgfsetlinewidth{1.003750pt}%
\definecolor{currentstroke}{rgb}{0.500000,0.500000,0.500000}%
\pgfsetstrokecolor{currentstroke}%
\pgfsetdash{{3.700000pt}{1.600000pt}}{0.000000pt}%
\pgfpathmoveto{\pgfqpoint{0.523611in}{2.083335in}}%
\pgfpathlineto{\pgfqpoint{0.745833in}{2.083335in}}%
\pgfusepath{stroke}%
\end{pgfscope}%
\begin{pgfscope}%
\definecolor{textcolor}{rgb}{0.000000,0.000000,0.000000}%
\pgfsetstrokecolor{textcolor}%
\pgfsetfillcolor{textcolor}%
\pgftext[x=0.834722in,y=2.044446in,left,base]{\color{textcolor}\rmfamily\fontsize{8.000000}{9.600000}\selectfont Linear Fit}%
\end{pgfscope}%
\begin{pgfscope}%
\pgfsetbuttcap%
\pgfsetroundjoin%
\pgfsetlinewidth{0.752812pt}%
\definecolor{currentstroke}{rgb}{0.000000,0.000000,0.000000}%
\pgfsetstrokecolor{currentstroke}%
\pgfsetdash{}{0pt}%
\pgfpathmoveto{\pgfqpoint{0.634722in}{1.867335in}}%
\pgfpathlineto{\pgfqpoint{0.634722in}{1.978446in}}%
\pgfusepath{stroke}%
\end{pgfscope}%
\begin{pgfscope}%
\pgfsetbuttcap%
\pgfsetroundjoin%
\definecolor{currentfill}{rgb}{0.000000,0.000000,0.000000}%
\pgfsetfillcolor{currentfill}%
\pgfsetfillopacity{0.000000}%
\pgfsetlinewidth{0.752812pt}%
\definecolor{currentstroke}{rgb}{0.000000,0.000000,0.000000}%
\pgfsetstrokecolor{currentstroke}%
\pgfsetdash{}{0pt}%
\pgfsys@defobject{currentmarker}{\pgfqpoint{-0.013889in}{-0.000000in}}{\pgfqpoint{0.013889in}{0.000000in}}{%
\pgfpathmoveto{\pgfqpoint{0.013889in}{-0.000000in}}%
\pgfpathlineto{\pgfqpoint{-0.013889in}{0.000000in}}%
\pgfusepath{stroke,fill}%
}%
\begin{pgfscope}%
\pgfsys@transformshift{0.634722in}{1.867335in}%
\pgfsys@useobject{currentmarker}{}%
\end{pgfscope}%
\end{pgfscope}%
\begin{pgfscope}%
\pgfsetbuttcap%
\pgfsetroundjoin%
\definecolor{currentfill}{rgb}{0.000000,0.000000,0.000000}%
\pgfsetfillcolor{currentfill}%
\pgfsetfillopacity{0.000000}%
\pgfsetlinewidth{0.752812pt}%
\definecolor{currentstroke}{rgb}{0.000000,0.000000,0.000000}%
\pgfsetstrokecolor{currentstroke}%
\pgfsetdash{}{0pt}%
\pgfsys@defobject{currentmarker}{\pgfqpoint{-0.013889in}{-0.000000in}}{\pgfqpoint{0.013889in}{0.000000in}}{%
\pgfpathmoveto{\pgfqpoint{0.013889in}{-0.000000in}}%
\pgfpathlineto{\pgfqpoint{-0.013889in}{0.000000in}}%
\pgfusepath{stroke,fill}%
}%
\begin{pgfscope}%
\pgfsys@transformshift{0.634722in}{1.978446in}%
\pgfsys@useobject{currentmarker}{}%
\end{pgfscope}%
\end{pgfscope}%
\begin{pgfscope}%
\pgfsetbuttcap%
\pgfsetroundjoin%
\definecolor{currentfill}{rgb}{0.000000,0.000000,0.000000}%
\pgfsetfillcolor{currentfill}%
\pgfsetlinewidth{1.003750pt}%
\definecolor{currentstroke}{rgb}{0.000000,0.000000,0.000000}%
\pgfsetstrokecolor{currentstroke}%
\pgfsetdash{}{0pt}%
\pgfsys@defobject{currentmarker}{\pgfqpoint{-0.010417in}{-0.010417in}}{\pgfqpoint{0.010417in}{0.010417in}}{%
\pgfpathmoveto{\pgfqpoint{0.000000in}{-0.010417in}}%
\pgfpathcurveto{\pgfqpoint{0.002763in}{-0.010417in}}{\pgfqpoint{0.005412in}{-0.009319in}}{\pgfqpoint{0.007366in}{-0.007366in}}%
\pgfpathcurveto{\pgfqpoint{0.009319in}{-0.005412in}}{\pgfqpoint{0.010417in}{-0.002763in}}{\pgfqpoint{0.010417in}{0.000000in}}%
\pgfpathcurveto{\pgfqpoint{0.010417in}{0.002763in}}{\pgfqpoint{0.009319in}{0.005412in}}{\pgfqpoint{0.007366in}{0.007366in}}%
\pgfpathcurveto{\pgfqpoint{0.005412in}{0.009319in}}{\pgfqpoint{0.002763in}{0.010417in}}{\pgfqpoint{0.000000in}{0.010417in}}%
\pgfpathcurveto{\pgfqpoint{-0.002763in}{0.010417in}}{\pgfqpoint{-0.005412in}{0.009319in}}{\pgfqpoint{-0.007366in}{0.007366in}}%
\pgfpathcurveto{\pgfqpoint{-0.009319in}{0.005412in}}{\pgfqpoint{-0.010417in}{0.002763in}}{\pgfqpoint{-0.010417in}{0.000000in}}%
\pgfpathcurveto{\pgfqpoint{-0.010417in}{-0.002763in}}{\pgfqpoint{-0.009319in}{-0.005412in}}{\pgfqpoint{-0.007366in}{-0.007366in}}%
\pgfpathcurveto{\pgfqpoint{-0.005412in}{-0.009319in}}{\pgfqpoint{-0.002763in}{-0.010417in}}{\pgfqpoint{0.000000in}{-0.010417in}}%
\pgfpathclose%
\pgfusepath{stroke,fill}%
}%
\begin{pgfscope}%
\pgfsys@transformshift{0.634722in}{1.922891in}%
\pgfsys@useobject{currentmarker}{}%
\end{pgfscope}%
\end{pgfscope}%
\begin{pgfscope}%
\definecolor{textcolor}{rgb}{0.000000,0.000000,0.000000}%
\pgfsetstrokecolor{textcolor}%
\pgfsetfillcolor{textcolor}%
\pgftext[x=0.834722in,y=1.884002in,left,base]{\color{textcolor}\rmfamily\fontsize{8.000000}{9.600000}\selectfont Measurement Deviation [\(\displaystyle \upmu\)A]}%
\end{pgfscope}%
\end{pgfpicture}%
\makeatother%
\endgroup%

%% file: figs/sample_power.pgf
\begingroup%
\makeatletter%
\begin{pgfpicture}%
\pgfpathrectangle{\pgfpointorigin}{\pgfqpoint{3.193753in}{1.805557in}}%
\pgfusepath{use as bounding box, clip}%
\begin{pgfscope}%
\pgfsetbuttcap%
\pgfsetmiterjoin%
\definecolor{currentfill}{rgb}{1.000000,1.000000,1.000000}%
\pgfsetfillcolor{currentfill}%
\pgfsetlinewidth{0.000000pt}%
\definecolor{currentstroke}{rgb}{1.000000,1.000000,1.000000}%
\pgfsetstrokecolor{currentstroke}%
\pgfsetdash{}{0pt}%
\pgfpathmoveto{\pgfqpoint{0.000000in}{0.000000in}}%
\pgfpathlineto{\pgfqpoint{3.193753in}{0.000000in}}%
\pgfpathlineto{\pgfqpoint{3.193753in}{1.805557in}}%
\pgfpathlineto{\pgfqpoint{0.000000in}{1.805557in}}%
\pgfpathclose%
\pgfusepath{fill}%
\end{pgfscope}%
\begin{pgfscope}%
\pgfsetbuttcap%
\pgfsetmiterjoin%
\definecolor{currentfill}{rgb}{1.000000,1.000000,1.000000}%
\pgfsetfillcolor{currentfill}%
\pgfsetlinewidth{0.000000pt}%
\definecolor{currentstroke}{rgb}{0.000000,0.000000,0.000000}%
\pgfsetstrokecolor{currentstroke}%
\pgfsetstrokeopacity{0.000000}%
\pgfsetdash{}{0pt}%
\pgfpathmoveto{\pgfqpoint{0.381944in}{0.425000in}}%
\pgfpathlineto{\pgfqpoint{3.167371in}{0.425000in}}%
\pgfpathlineto{\pgfqpoint{3.167371in}{1.805557in}}%
\pgfpathlineto{\pgfqpoint{0.381944in}{1.805557in}}%
\pgfpathclose%
\pgfusepath{fill}%
\end{pgfscope}%
\begin{pgfscope}%
\pgfsetbuttcap%
\pgfsetroundjoin%
\definecolor{currentfill}{rgb}{0.000000,0.000000,0.000000}%
\pgfsetfillcolor{currentfill}%
\pgfsetfillopacity{0.000000}%
\pgfsetlinewidth{0.803000pt}%
\definecolor{currentstroke}{rgb}{0.000000,0.000000,0.000000}%
\pgfsetstrokecolor{currentstroke}%
\pgfsetdash{}{0pt}%
\pgfsys@defobject{currentmarker}{\pgfqpoint{0.000000in}{-0.069444in}}{\pgfqpoint{0.000000in}{0.000000in}}{%
\pgfpathmoveto{\pgfqpoint{0.000000in}{0.000000in}}%
\pgfpathlineto{\pgfqpoint{0.000000in}{-0.069444in}}%
\pgfusepath{stroke,fill}%
}%
\begin{pgfscope}%
\pgfsys@transformshift{0.508554in}{0.425000in}%
\pgfsys@useobject{currentmarker}{}%
\end{pgfscope}%
\end{pgfscope}%
\begin{pgfscope}%
\definecolor{textcolor}{rgb}{0.000000,0.000000,0.000000}%
\pgfsetstrokecolor{textcolor}%
\pgfsetfillcolor{textcolor}%
\pgftext[x=0.508554in,y=0.306944in,,top]{\color{textcolor}\rmfamily\fontsize{10.000000}{12.000000}\selectfont 1}%
\end{pgfscope}%
\begin{pgfscope}%
\pgfsetbuttcap%
\pgfsetroundjoin%
\definecolor{currentfill}{rgb}{0.000000,0.000000,0.000000}%
\pgfsetfillcolor{currentfill}%
\pgfsetfillopacity{0.000000}%
\pgfsetlinewidth{0.803000pt}%
\definecolor{currentstroke}{rgb}{0.000000,0.000000,0.000000}%
\pgfsetstrokecolor{currentstroke}%
\pgfsetdash{}{0pt}%
\pgfsys@defobject{currentmarker}{\pgfqpoint{0.000000in}{-0.069444in}}{\pgfqpoint{0.000000in}{0.000000in}}{%
\pgfpathmoveto{\pgfqpoint{0.000000in}{0.000000in}}%
\pgfpathlineto{\pgfqpoint{0.000000in}{-0.069444in}}%
\pgfusepath{stroke,fill}%
}%
\begin{pgfscope}%
\pgfsys@transformshift{0.870298in}{0.425000in}%
\pgfsys@useobject{currentmarker}{}%
\end{pgfscope}%
\end{pgfscope}%
\begin{pgfscope}%
\definecolor{textcolor}{rgb}{0.000000,0.000000,0.000000}%
\pgfsetstrokecolor{textcolor}%
\pgfsetfillcolor{textcolor}%
\pgftext[x=0.870298in,y=0.306944in,,top]{\color{textcolor}\rmfamily\fontsize{10.000000}{12.000000}\selectfont 4}%
\end{pgfscope}%
\begin{pgfscope}%
\pgfsetbuttcap%
\pgfsetroundjoin%
\definecolor{currentfill}{rgb}{0.000000,0.000000,0.000000}%
\pgfsetfillcolor{currentfill}%
\pgfsetfillopacity{0.000000}%
\pgfsetlinewidth{0.803000pt}%
\definecolor{currentstroke}{rgb}{0.000000,0.000000,0.000000}%
\pgfsetstrokecolor{currentstroke}%
\pgfsetdash{}{0pt}%
\pgfsys@defobject{currentmarker}{\pgfqpoint{0.000000in}{-0.069444in}}{\pgfqpoint{0.000000in}{0.000000in}}{%
\pgfpathmoveto{\pgfqpoint{0.000000in}{0.000000in}}%
\pgfpathlineto{\pgfqpoint{0.000000in}{-0.069444in}}%
\pgfusepath{stroke,fill}%
}%
\begin{pgfscope}%
\pgfsys@transformshift{1.232042in}{0.425000in}%
\pgfsys@useobject{currentmarker}{}%
\end{pgfscope}%
\end{pgfscope}%
\begin{pgfscope}%
\definecolor{textcolor}{rgb}{0.000000,0.000000,0.000000}%
\pgfsetstrokecolor{textcolor}%
\pgfsetfillcolor{textcolor}%
\pgftext[x=1.232042in,y=0.306944in,,top]{\color{textcolor}\rmfamily\fontsize{10.000000}{12.000000}\selectfont 16}%
\end{pgfscope}%
\begin{pgfscope}%
\pgfsetbuttcap%
\pgfsetroundjoin%
\definecolor{currentfill}{rgb}{0.000000,0.000000,0.000000}%
\pgfsetfillcolor{currentfill}%
\pgfsetfillopacity{0.000000}%
\pgfsetlinewidth{0.803000pt}%
\definecolor{currentstroke}{rgb}{0.000000,0.000000,0.000000}%
\pgfsetstrokecolor{currentstroke}%
\pgfsetdash{}{0pt}%
\pgfsys@defobject{currentmarker}{\pgfqpoint{0.000000in}{-0.069444in}}{\pgfqpoint{0.000000in}{0.000000in}}{%
\pgfpathmoveto{\pgfqpoint{0.000000in}{0.000000in}}%
\pgfpathlineto{\pgfqpoint{0.000000in}{-0.069444in}}%
\pgfusepath{stroke,fill}%
}%
\begin{pgfscope}%
\pgfsys@transformshift{1.593786in}{0.425000in}%
\pgfsys@useobject{currentmarker}{}%
\end{pgfscope}%
\end{pgfscope}%
\begin{pgfscope}%
\definecolor{textcolor}{rgb}{0.000000,0.000000,0.000000}%
\pgfsetstrokecolor{textcolor}%
\pgfsetfillcolor{textcolor}%
\pgftext[x=1.593786in,y=0.306944in,,top]{\color{textcolor}\rmfamily\fontsize{10.000000}{12.000000}\selectfont 64}%
\end{pgfscope}%
\begin{pgfscope}%
\pgfsetbuttcap%
\pgfsetroundjoin%
\definecolor{currentfill}{rgb}{0.000000,0.000000,0.000000}%
\pgfsetfillcolor{currentfill}%
\pgfsetfillopacity{0.000000}%
\pgfsetlinewidth{0.803000pt}%
\definecolor{currentstroke}{rgb}{0.000000,0.000000,0.000000}%
\pgfsetstrokecolor{currentstroke}%
\pgfsetdash{}{0pt}%
\pgfsys@defobject{currentmarker}{\pgfqpoint{0.000000in}{-0.069444in}}{\pgfqpoint{0.000000in}{0.000000in}}{%
\pgfpathmoveto{\pgfqpoint{0.000000in}{0.000000in}}%
\pgfpathlineto{\pgfqpoint{0.000000in}{-0.069444in}}%
\pgfusepath{stroke,fill}%
}%
\begin{pgfscope}%
\pgfsys@transformshift{1.955529in}{0.425000in}%
\pgfsys@useobject{currentmarker}{}%
\end{pgfscope}%
\end{pgfscope}%
\begin{pgfscope}%
\definecolor{textcolor}{rgb}{0.000000,0.000000,0.000000}%
\pgfsetstrokecolor{textcolor}%
\pgfsetfillcolor{textcolor}%
\pgftext[x=1.955529in,y=0.306944in,,top]{\color{textcolor}\rmfamily\fontsize{10.000000}{12.000000}\selectfont 128}%
\end{pgfscope}%
\begin{pgfscope}%
\pgfsetbuttcap%
\pgfsetroundjoin%
\definecolor{currentfill}{rgb}{0.000000,0.000000,0.000000}%
\pgfsetfillcolor{currentfill}%
\pgfsetfillopacity{0.000000}%
\pgfsetlinewidth{0.803000pt}%
\definecolor{currentstroke}{rgb}{0.000000,0.000000,0.000000}%
\pgfsetstrokecolor{currentstroke}%
\pgfsetdash{}{0pt}%
\pgfsys@defobject{currentmarker}{\pgfqpoint{0.000000in}{-0.069444in}}{\pgfqpoint{0.000000in}{0.000000in}}{%
\pgfpathmoveto{\pgfqpoint{0.000000in}{0.000000in}}%
\pgfpathlineto{\pgfqpoint{0.000000in}{-0.069444in}}%
\pgfusepath{stroke,fill}%
}%
\begin{pgfscope}%
\pgfsys@transformshift{2.317273in}{0.425000in}%
\pgfsys@useobject{currentmarker}{}%
\end{pgfscope}%
\end{pgfscope}%
\begin{pgfscope}%
\definecolor{textcolor}{rgb}{0.000000,0.000000,0.000000}%
\pgfsetstrokecolor{textcolor}%
\pgfsetfillcolor{textcolor}%
\pgftext[x=2.317273in,y=0.306944in,,top]{\color{textcolor}\rmfamily\fontsize{10.000000}{12.000000}\selectfont 256}%
\end{pgfscope}%
\begin{pgfscope}%
\pgfsetbuttcap%
\pgfsetroundjoin%
\definecolor{currentfill}{rgb}{0.000000,0.000000,0.000000}%
\pgfsetfillcolor{currentfill}%
\pgfsetfillopacity{0.000000}%
\pgfsetlinewidth{0.803000pt}%
\definecolor{currentstroke}{rgb}{0.000000,0.000000,0.000000}%
\pgfsetstrokecolor{currentstroke}%
\pgfsetdash{}{0pt}%
\pgfsys@defobject{currentmarker}{\pgfqpoint{0.000000in}{-0.069444in}}{\pgfqpoint{0.000000in}{0.000000in}}{%
\pgfpathmoveto{\pgfqpoint{0.000000in}{0.000000in}}%
\pgfpathlineto{\pgfqpoint{0.000000in}{-0.069444in}}%
\pgfusepath{stroke,fill}%
}%
\begin{pgfscope}%
\pgfsys@transformshift{2.679017in}{0.425000in}%
\pgfsys@useobject{currentmarker}{}%
\end{pgfscope}%
\end{pgfscope}%
\begin{pgfscope}%
\definecolor{textcolor}{rgb}{0.000000,0.000000,0.000000}%
\pgfsetstrokecolor{textcolor}%
\pgfsetfillcolor{textcolor}%
\pgftext[x=2.679017in,y=0.306944in,,top]{\color{textcolor}\rmfamily\fontsize{10.000000}{12.000000}\selectfont 512}%
\end{pgfscope}%
\begin{pgfscope}%
\pgfsetbuttcap%
\pgfsetroundjoin%
\definecolor{currentfill}{rgb}{0.000000,0.000000,0.000000}%
\pgfsetfillcolor{currentfill}%
\pgfsetfillopacity{0.000000}%
\pgfsetlinewidth{0.803000pt}%
\definecolor{currentstroke}{rgb}{0.000000,0.000000,0.000000}%
\pgfsetstrokecolor{currentstroke}%
\pgfsetdash{}{0pt}%
\pgfsys@defobject{currentmarker}{\pgfqpoint{0.000000in}{-0.069444in}}{\pgfqpoint{0.000000in}{0.000000in}}{%
\pgfpathmoveto{\pgfqpoint{0.000000in}{0.000000in}}%
\pgfpathlineto{\pgfqpoint{0.000000in}{-0.069444in}}%
\pgfusepath{stroke,fill}%
}%
\begin{pgfscope}%
\pgfsys@transformshift{3.040761in}{0.425000in}%
\pgfsys@useobject{currentmarker}{}%
\end{pgfscope}%
\end{pgfscope}%
\begin{pgfscope}%
\definecolor{textcolor}{rgb}{0.000000,0.000000,0.000000}%
\pgfsetstrokecolor{textcolor}%
\pgfsetfillcolor{textcolor}%
\pgftext[x=3.040761in,y=0.306944in,,top]{\color{textcolor}\rmfamily\fontsize{10.000000}{12.000000}\selectfont 1024}%
\end{pgfscope}%
\begin{pgfscope}%
\definecolor{textcolor}{rgb}{0.000000,0.000000,0.000000}%
\pgfsetstrokecolor{textcolor}%
\pgfsetfillcolor{textcolor}%
\pgftext[x=1.774658in,y=0.128055in,,top]{\color{textcolor}\rmfamily\fontsize{10.000000}{12.000000}\selectfont \# of Averaged Samples}%
\end{pgfscope}%
\begin{pgfscope}%
\pgfsetbuttcap%
\pgfsetroundjoin%
\definecolor{currentfill}{rgb}{0.000000,0.000000,0.000000}%
\pgfsetfillcolor{currentfill}%
\pgfsetfillopacity{0.000000}%
\pgfsetlinewidth{0.803000pt}%
\definecolor{currentstroke}{rgb}{0.000000,0.000000,0.000000}%
\pgfsetstrokecolor{currentstroke}%
\pgfsetdash{}{0pt}%
\pgfsys@defobject{currentmarker}{\pgfqpoint{-0.069444in}{0.000000in}}{\pgfqpoint{0.000000in}{0.000000in}}{%
\pgfpathmoveto{\pgfqpoint{0.000000in}{0.000000in}}%
\pgfpathlineto{\pgfqpoint{-0.069444in}{0.000000in}}%
\pgfusepath{stroke,fill}%
}%
\begin{pgfscope}%
\pgfsys@transformshift{0.381944in}{0.425000in}%
\pgfsys@useobject{currentmarker}{}%
\end{pgfscope}%
\end{pgfscope}%
\begin{pgfscope}%
\definecolor{textcolor}{rgb}{0.000000,0.000000,0.000000}%
\pgfsetstrokecolor{textcolor}%
\pgfsetfillcolor{textcolor}%
\pgftext[x=0.194444in,y=0.376805in,left,base]{\color{textcolor}\rmfamily\fontsize{10.000000}{12.000000}\selectfont \(\displaystyle 1\)}%
\end{pgfscope}%
\begin{pgfscope}%
\pgfsetbuttcap%
\pgfsetroundjoin%
\definecolor{currentfill}{rgb}{0.000000,0.000000,0.000000}%
\pgfsetfillcolor{currentfill}%
\pgfsetfillopacity{0.000000}%
\pgfsetlinewidth{0.803000pt}%
\definecolor{currentstroke}{rgb}{0.000000,0.000000,0.000000}%
\pgfsetstrokecolor{currentstroke}%
\pgfsetdash{}{0pt}%
\pgfsys@defobject{currentmarker}{\pgfqpoint{-0.069444in}{0.000000in}}{\pgfqpoint{0.000000in}{0.000000in}}{%
\pgfpathmoveto{\pgfqpoint{0.000000in}{0.000000in}}%
\pgfpathlineto{\pgfqpoint{-0.069444in}{0.000000in}}%
\pgfusepath{stroke,fill}%
}%
\begin{pgfscope}%
\pgfsys@transformshift{0.381944in}{0.819445in}%
\pgfsys@useobject{currentmarker}{}%
\end{pgfscope}%
\end{pgfscope}%
\begin{pgfscope}%
\definecolor{textcolor}{rgb}{0.000000,0.000000,0.000000}%
\pgfsetstrokecolor{textcolor}%
\pgfsetfillcolor{textcolor}%
\pgftext[x=0.194444in,y=0.771250in,left,base]{\color{textcolor}\rmfamily\fontsize{10.000000}{12.000000}\selectfont \(\displaystyle 2\)}%
\end{pgfscope}%
\begin{pgfscope}%
\pgfsetbuttcap%
\pgfsetroundjoin%
\definecolor{currentfill}{rgb}{0.000000,0.000000,0.000000}%
\pgfsetfillcolor{currentfill}%
\pgfsetfillopacity{0.000000}%
\pgfsetlinewidth{0.803000pt}%
\definecolor{currentstroke}{rgb}{0.000000,0.000000,0.000000}%
\pgfsetstrokecolor{currentstroke}%
\pgfsetdash{}{0pt}%
\pgfsys@defobject{currentmarker}{\pgfqpoint{-0.069444in}{0.000000in}}{\pgfqpoint{0.000000in}{0.000000in}}{%
\pgfpathmoveto{\pgfqpoint{0.000000in}{0.000000in}}%
\pgfpathlineto{\pgfqpoint{-0.069444in}{0.000000in}}%
\pgfusepath{stroke,fill}%
}%
\begin{pgfscope}%
\pgfsys@transformshift{0.381944in}{1.213890in}%
\pgfsys@useobject{currentmarker}{}%
\end{pgfscope}%
\end{pgfscope}%
\begin{pgfscope}%
\definecolor{textcolor}{rgb}{0.000000,0.000000,0.000000}%
\pgfsetstrokecolor{textcolor}%
\pgfsetfillcolor{textcolor}%
\pgftext[x=0.194444in,y=1.165695in,left,base]{\color{textcolor}\rmfamily\fontsize{10.000000}{12.000000}\selectfont \(\displaystyle 3\)}%
\end{pgfscope}%
\begin{pgfscope}%
\pgfsetbuttcap%
\pgfsetroundjoin%
\definecolor{currentfill}{rgb}{0.000000,0.000000,0.000000}%
\pgfsetfillcolor{currentfill}%
\pgfsetfillopacity{0.000000}%
\pgfsetlinewidth{0.803000pt}%
\definecolor{currentstroke}{rgb}{0.000000,0.000000,0.000000}%
\pgfsetstrokecolor{currentstroke}%
\pgfsetdash{}{0pt}%
\pgfsys@defobject{currentmarker}{\pgfqpoint{-0.069444in}{0.000000in}}{\pgfqpoint{0.000000in}{0.000000in}}{%
\pgfpathmoveto{\pgfqpoint{0.000000in}{0.000000in}}%
\pgfpathlineto{\pgfqpoint{-0.069444in}{0.000000in}}%
\pgfusepath{stroke,fill}%
}%
\begin{pgfscope}%
\pgfsys@transformshift{0.381944in}{1.608335in}%
\pgfsys@useobject{currentmarker}{}%
\end{pgfscope}%
\end{pgfscope}%
\begin{pgfscope}%
\definecolor{textcolor}{rgb}{0.000000,0.000000,0.000000}%
\pgfsetstrokecolor{textcolor}%
\pgfsetfillcolor{textcolor}%
\pgftext[x=0.194444in,y=1.560140in,left,base]{\color{textcolor}\rmfamily\fontsize{10.000000}{12.000000}\selectfont \(\displaystyle 4\)}%
\end{pgfscope}%
\begin{pgfscope}%
\definecolor{textcolor}{rgb}{0.000000,0.000000,0.000000}%
\pgfsetstrokecolor{textcolor}%
\pgfsetfillcolor{textcolor}%
\pgftext[x=0.138889in,y=1.115278in,,bottom,rotate=90.000000]{\color{textcolor}\rmfamily\fontsize{10.000000}{12.000000}\selectfont Power [mW]}%
\end{pgfscope}%
\begin{pgfscope}%
\pgfpathrectangle{\pgfqpoint{0.381944in}{0.425000in}}{\pgfqpoint{2.785427in}{1.380557in}}%
\pgfusepath{clip}%
\pgfsetbuttcap%
\pgfsetroundjoin%
\definecolor{currentfill}{rgb}{0.000000,0.000000,0.000000}%
\pgfsetfillcolor{currentfill}%
\pgfsetfillopacity{0.000000}%
\pgfsetlinewidth{1.003750pt}%
\definecolor{currentstroke}{rgb}{0.000000,0.000000,0.000000}%
\pgfsetstrokecolor{currentstroke}%
\pgfsetdash{}{0pt}%
\pgfsys@defobject{currentmarker}{\pgfqpoint{-0.034722in}{-0.034722in}}{\pgfqpoint{0.034722in}{0.034722in}}{%
\pgfpathmoveto{\pgfqpoint{-0.034722in}{-0.034722in}}%
\pgfpathlineto{\pgfqpoint{0.034722in}{0.034722in}}%
\pgfpathmoveto{\pgfqpoint{-0.034722in}{0.034722in}}%
\pgfpathlineto{\pgfqpoint{0.034722in}{-0.034722in}}%
\pgfusepath{stroke,fill}%
}%
\begin{pgfscope}%
\pgfsys@transformshift{0.508554in}{1.678024in}%
\pgfsys@useobject{currentmarker}{}%
\end{pgfscope}%
\begin{pgfscope}%
\pgfsys@transformshift{0.870298in}{0.725236in}%
\pgfsys@useobject{currentmarker}{}%
\end{pgfscope}%
\begin{pgfscope}%
\pgfsys@transformshift{1.232042in}{0.510065in}%
\pgfsys@useobject{currentmarker}{}%
\end{pgfscope}%
\begin{pgfscope}%
\pgfsys@transformshift{1.593786in}{0.459903in}%
\pgfsys@useobject{currentmarker}{}%
\end{pgfscope}%
\begin{pgfscope}%
\pgfsys@transformshift{1.955529in}{0.451829in}%
\pgfsys@useobject{currentmarker}{}%
\end{pgfscope}%
\begin{pgfscope}%
\pgfsys@transformshift{2.317273in}{0.448004in}%
\pgfsys@useobject{currentmarker}{}%
\end{pgfscope}%
\begin{pgfscope}%
\pgfsys@transformshift{2.679017in}{0.445584in}%
\pgfsys@useobject{currentmarker}{}%
\end{pgfscope}%
\begin{pgfscope}%
\pgfsys@transformshift{3.040761in}{0.444558in}%
\pgfsys@useobject{currentmarker}{}%
\end{pgfscope}%
\end{pgfscope}%
\begin{pgfscope}%
\pgfpathrectangle{\pgfqpoint{0.381944in}{0.425000in}}{\pgfqpoint{2.785427in}{1.380557in}}%
\pgfusepath{clip}%
\pgfsetbuttcap%
\pgfsetroundjoin%
\definecolor{currentfill}{rgb}{0.000000,0.000000,0.000000}%
\pgfsetfillcolor{currentfill}%
\pgfsetfillopacity{0.000000}%
\pgfsetlinewidth{1.003750pt}%
\definecolor{currentstroke}{rgb}{0.000000,0.000000,0.000000}%
\pgfsetstrokecolor{currentstroke}%
\pgfsetdash{}{0pt}%
\pgfsys@defobject{currentmarker}{\pgfqpoint{-0.034722in}{-0.034722in}}{\pgfqpoint{0.034722in}{0.034722in}}{%
\pgfpathmoveto{\pgfqpoint{-0.034722in}{0.000000in}}%
\pgfpathlineto{\pgfqpoint{0.034722in}{0.000000in}}%
\pgfpathmoveto{\pgfqpoint{0.000000in}{-0.034722in}}%
\pgfpathlineto{\pgfqpoint{0.000000in}{0.034722in}}%
\pgfusepath{stroke,fill}%
}%
\begin{pgfscope}%
\pgfsys@transformshift{0.508554in}{0.747672in}%
\pgfsys@useobject{currentmarker}{}%
\end{pgfscope}%
\begin{pgfscope}%
\pgfsys@transformshift{0.870298in}{0.520020in}%
\pgfsys@useobject{currentmarker}{}%
\end{pgfscope}%
\begin{pgfscope}%
\pgfsys@transformshift{1.232042in}{0.468515in}%
\pgfsys@useobject{currentmarker}{}%
\end{pgfscope}%
\begin{pgfscope}%
\pgfsys@transformshift{1.593786in}{0.455137in}%
\pgfsys@useobject{currentmarker}{}%
\end{pgfscope}%
\begin{pgfscope}%
\pgfsys@transformshift{1.955529in}{0.452808in}%
\pgfsys@useobject{currentmarker}{}%
\end{pgfscope}%
\begin{pgfscope}%
\pgfsys@transformshift{2.317273in}{0.451922in}%
\pgfsys@useobject{currentmarker}{}%
\end{pgfscope}%
\begin{pgfscope}%
\pgfsys@transformshift{2.679017in}{0.451350in}%
\pgfsys@useobject{currentmarker}{}%
\end{pgfscope}%
\begin{pgfscope}%
\pgfsys@transformshift{3.040761in}{0.451156in}%
\pgfsys@useobject{currentmarker}{}%
\end{pgfscope}%
\end{pgfscope}%
\begin{pgfscope}%
\pgfpathrectangle{\pgfqpoint{0.381944in}{0.425000in}}{\pgfqpoint{2.785427in}{1.380557in}}%
\pgfusepath{clip}%
\pgfsetbuttcap%
\pgfsetroundjoin%
\definecolor{currentfill}{rgb}{0.000000,0.000000,0.000000}%
\pgfsetfillcolor{currentfill}%
\pgfsetfillopacity{0.000000}%
\pgfsetlinewidth{1.003750pt}%
\definecolor{currentstroke}{rgb}{0.000000,0.000000,0.000000}%
\pgfsetstrokecolor{currentstroke}%
\pgfsetdash{}{0pt}%
\pgfsys@defobject{currentmarker}{\pgfqpoint{-0.034722in}{-0.034722in}}{\pgfqpoint{0.034722in}{0.034722in}}{%
\pgfpathmoveto{\pgfqpoint{0.000000in}{-0.034722in}}%
\pgfpathcurveto{\pgfqpoint{0.009208in}{-0.034722in}}{\pgfqpoint{0.018041in}{-0.031064in}}{\pgfqpoint{0.024552in}{-0.024552in}}%
\pgfpathcurveto{\pgfqpoint{0.031064in}{-0.018041in}}{\pgfqpoint{0.034722in}{-0.009208in}}{\pgfqpoint{0.034722in}{0.000000in}}%
\pgfpathcurveto{\pgfqpoint{0.034722in}{0.009208in}}{\pgfqpoint{0.031064in}{0.018041in}}{\pgfqpoint{0.024552in}{0.024552in}}%
\pgfpathcurveto{\pgfqpoint{0.018041in}{0.031064in}}{\pgfqpoint{0.009208in}{0.034722in}}{\pgfqpoint{0.000000in}{0.034722in}}%
\pgfpathcurveto{\pgfqpoint{-0.009208in}{0.034722in}}{\pgfqpoint{-0.018041in}{0.031064in}}{\pgfqpoint{-0.024552in}{0.024552in}}%
\pgfpathcurveto{\pgfqpoint{-0.031064in}{0.018041in}}{\pgfqpoint{-0.034722in}{0.009208in}}{\pgfqpoint{-0.034722in}{0.000000in}}%
\pgfpathcurveto{\pgfqpoint{-0.034722in}{-0.009208in}}{\pgfqpoint{-0.031064in}{-0.018041in}}{\pgfqpoint{-0.024552in}{-0.024552in}}%
\pgfpathcurveto{\pgfqpoint{-0.018041in}{-0.031064in}}{\pgfqpoint{-0.009208in}{-0.034722in}}{\pgfqpoint{0.000000in}{-0.034722in}}%
\pgfpathclose%
\pgfusepath{stroke,fill}%
}%
\begin{pgfscope}%
\pgfsys@transformshift{0.508554in}{0.528960in}%
\pgfsys@useobject{currentmarker}{}%
\end{pgfscope}%
\begin{pgfscope}%
\pgfsys@transformshift{0.870298in}{0.471270in}%
\pgfsys@useobject{currentmarker}{}%
\end{pgfscope}%
\begin{pgfscope}%
\pgfsys@transformshift{1.232042in}{0.457093in}%
\pgfsys@useobject{currentmarker}{}%
\end{pgfscope}%
\begin{pgfscope}%
\pgfsys@transformshift{1.593786in}{0.453290in}%
\pgfsys@useobject{currentmarker}{}%
\end{pgfscope}%
\begin{pgfscope}%
\pgfsys@transformshift{1.955529in}{0.452792in}%
\pgfsys@useobject{currentmarker}{}%
\end{pgfscope}%
\begin{pgfscope}%
\pgfsys@transformshift{2.317273in}{0.452502in}%
\pgfsys@useobject{currentmarker}{}%
\end{pgfscope}%
\begin{pgfscope}%
\pgfsys@transformshift{2.679017in}{0.452369in}%
\pgfsys@useobject{currentmarker}{}%
\end{pgfscope}%
\begin{pgfscope}%
\pgfsys@transformshift{3.040761in}{0.452287in}%
\pgfsys@useobject{currentmarker}{}%
\end{pgfscope}%
\end{pgfscope}%
\begin{pgfscope}%
\pgfpathrectangle{\pgfqpoint{0.381944in}{0.425000in}}{\pgfqpoint{2.785427in}{1.380557in}}%
\pgfusepath{clip}%
\pgfsetbuttcap%
\pgfsetmiterjoin%
\definecolor{currentfill}{rgb}{0.000000,0.000000,0.000000}%
\pgfsetfillcolor{currentfill}%
\pgfsetfillopacity{0.000000}%
\pgfsetlinewidth{1.003750pt}%
\definecolor{currentstroke}{rgb}{0.000000,0.000000,0.000000}%
\pgfsetstrokecolor{currentstroke}%
\pgfsetdash{}{0pt}%
\pgfsys@defobject{currentmarker}{\pgfqpoint{-0.034722in}{-0.034722in}}{\pgfqpoint{0.034722in}{0.034722in}}{%
\pgfpathmoveto{\pgfqpoint{0.000000in}{0.034722in}}%
\pgfpathlineto{\pgfqpoint{-0.034722in}{-0.034722in}}%
\pgfpathlineto{\pgfqpoint{0.034722in}{-0.034722in}}%
\pgfpathclose%
\pgfusepath{stroke,fill}%
}%
\begin{pgfscope}%
\pgfsys@transformshift{0.508554in}{0.472049in}%
\pgfsys@useobject{currentmarker}{}%
\end{pgfscope}%
\begin{pgfscope}%
\pgfsys@transformshift{0.870298in}{0.457395in}%
\pgfsys@useobject{currentmarker}{}%
\end{pgfscope}%
\begin{pgfscope}%
\pgfsys@transformshift{1.232042in}{0.453748in}%
\pgfsys@useobject{currentmarker}{}%
\end{pgfscope}%
\begin{pgfscope}%
\pgfsys@transformshift{1.593786in}{0.452846in}%
\pgfsys@useobject{currentmarker}{}%
\end{pgfscope}%
\begin{pgfscope}%
\pgfsys@transformshift{1.955529in}{0.452670in}%
\pgfsys@useobject{currentmarker}{}%
\end{pgfscope}%
\begin{pgfscope}%
\pgfsys@transformshift{2.317273in}{0.452606in}%
\pgfsys@useobject{currentmarker}{}%
\end{pgfscope}%
\begin{pgfscope}%
\pgfsys@transformshift{2.679017in}{0.452597in}%
\pgfsys@useobject{currentmarker}{}%
\end{pgfscope}%
\begin{pgfscope}%
\pgfsys@transformshift{3.040761in}{0.452622in}%
\pgfsys@useobject{currentmarker}{}%
\end{pgfscope}%
\end{pgfscope}%
\begin{pgfscope}%
\pgfsetrectcap%
\pgfsetmiterjoin%
\pgfsetlinewidth{0.803000pt}%
\definecolor{currentstroke}{rgb}{0.000000,0.000000,0.000000}%
\pgfsetstrokecolor{currentstroke}%
\pgfsetdash{}{0pt}%
\pgfpathmoveto{\pgfqpoint{0.381944in}{0.425000in}}%
\pgfpathlineto{\pgfqpoint{0.381944in}{1.805557in}}%
\pgfusepath{stroke}%
\end{pgfscope}%
\begin{pgfscope}%
\pgfsetrectcap%
\pgfsetmiterjoin%
\pgfsetlinewidth{0.803000pt}%
\definecolor{currentstroke}{rgb}{0.000000,0.000000,0.000000}%
\pgfsetstrokecolor{currentstroke}%
\pgfsetdash{}{0pt}%
\pgfpathmoveto{\pgfqpoint{3.167371in}{0.425000in}}%
\pgfpathlineto{\pgfqpoint{3.167371in}{1.805557in}}%
\pgfusepath{stroke}%
\end{pgfscope}%
\begin{pgfscope}%
\pgfsetrectcap%
\pgfsetmiterjoin%
\pgfsetlinewidth{0.803000pt}%
\definecolor{currentstroke}{rgb}{0.000000,0.000000,0.000000}%
\pgfsetstrokecolor{currentstroke}%
\pgfsetdash{}{0pt}%
\pgfpathmoveto{\pgfqpoint{0.381944in}{0.425000in}}%
\pgfpathlineto{\pgfqpoint{3.167371in}{0.425000in}}%
\pgfusepath{stroke}%
\end{pgfscope}%
\begin{pgfscope}%
\pgfsetrectcap%
\pgfsetmiterjoin%
\pgfsetlinewidth{0.803000pt}%
\definecolor{currentstroke}{rgb}{0.000000,0.000000,0.000000}%
\pgfsetstrokecolor{currentstroke}%
\pgfsetdash{}{0pt}%
\pgfpathmoveto{\pgfqpoint{0.381944in}{1.805557in}}%
\pgfpathlineto{\pgfqpoint{3.167371in}{1.805557in}}%
\pgfusepath{stroke}%
\end{pgfscope}%
\begin{pgfscope}%
\definecolor{textcolor}{rgb}{0.000000,0.000000,0.000000}%
\pgfsetstrokecolor{textcolor}%
\pgfsetfillcolor{textcolor}%
\pgftext[x=2.036260in,y=1.609168in,left,base]{\color{textcolor}\rmfamily\fontsize{10.000000}{12.000000}\selectfont Conversion Time}%
\end{pgfscope}%
\begin{pgfscope}%
\pgfsetbuttcap%
\pgfsetroundjoin%
\definecolor{currentfill}{rgb}{0.000000,0.000000,0.000000}%
\pgfsetfillcolor{currentfill}%
\pgfsetfillopacity{0.000000}%
\pgfsetlinewidth{1.003750pt}%
\definecolor{currentstroke}{rgb}{0.000000,0.000000,0.000000}%
\pgfsetstrokecolor{currentstroke}%
\pgfsetdash{}{0pt}%
\pgfsys@defobject{currentmarker}{\pgfqpoint{-0.034722in}{-0.034722in}}{\pgfqpoint{0.034722in}{0.034722in}}{%
\pgfpathmoveto{\pgfqpoint{-0.034722in}{-0.034722in}}%
\pgfpathlineto{\pgfqpoint{0.034722in}{0.034722in}}%
\pgfpathmoveto{\pgfqpoint{-0.034722in}{0.034722in}}%
\pgfpathlineto{\pgfqpoint{0.034722in}{-0.034722in}}%
\pgfusepath{stroke,fill}%
}%
\begin{pgfscope}%
\pgfsys@transformshift{2.313649in}{1.487779in}%
\pgfsys@useobject{currentmarker}{}%
\end{pgfscope}%
\end{pgfscope}%
\begin{pgfscope}%
\definecolor{textcolor}{rgb}{0.000000,0.000000,0.000000}%
\pgfsetstrokecolor{textcolor}%
\pgfsetfillcolor{textcolor}%
\pgftext[x=2.513649in,y=1.448890in,left,base]{\color{textcolor}\rmfamily\fontsize{8.000000}{9.600000}\selectfont 140 µs}%
\end{pgfscope}%
\begin{pgfscope}%
\pgfsetbuttcap%
\pgfsetroundjoin%
\definecolor{currentfill}{rgb}{0.000000,0.000000,0.000000}%
\pgfsetfillcolor{currentfill}%
\pgfsetfillopacity{0.000000}%
\pgfsetlinewidth{1.003750pt}%
\definecolor{currentstroke}{rgb}{0.000000,0.000000,0.000000}%
\pgfsetstrokecolor{currentstroke}%
\pgfsetdash{}{0pt}%
\pgfsys@defobject{currentmarker}{\pgfqpoint{-0.034722in}{-0.034722in}}{\pgfqpoint{0.034722in}{0.034722in}}{%
\pgfpathmoveto{\pgfqpoint{-0.034722in}{0.000000in}}%
\pgfpathlineto{\pgfqpoint{0.034722in}{0.000000in}}%
\pgfpathmoveto{\pgfqpoint{0.000000in}{-0.034722in}}%
\pgfpathlineto{\pgfqpoint{0.000000in}{0.034722in}}%
\pgfusepath{stroke,fill}%
}%
\begin{pgfscope}%
\pgfsys@transformshift{2.313649in}{1.332891in}%
\pgfsys@useobject{currentmarker}{}%
\end{pgfscope}%
\end{pgfscope}%
\begin{pgfscope}%
\definecolor{textcolor}{rgb}{0.000000,0.000000,0.000000}%
\pgfsetstrokecolor{textcolor}%
\pgfsetfillcolor{textcolor}%
\pgftext[x=2.513649in,y=1.294002in,left,base]{\color{textcolor}\rmfamily\fontsize{8.000000}{9.600000}\selectfont 588 µs}%
\end{pgfscope}%
\begin{pgfscope}%
\pgfsetbuttcap%
\pgfsetroundjoin%
\definecolor{currentfill}{rgb}{0.000000,0.000000,0.000000}%
\pgfsetfillcolor{currentfill}%
\pgfsetfillopacity{0.000000}%
\pgfsetlinewidth{1.003750pt}%
\definecolor{currentstroke}{rgb}{0.000000,0.000000,0.000000}%
\pgfsetstrokecolor{currentstroke}%
\pgfsetdash{}{0pt}%
\pgfsys@defobject{currentmarker}{\pgfqpoint{-0.034722in}{-0.034722in}}{\pgfqpoint{0.034722in}{0.034722in}}{%
\pgfpathmoveto{\pgfqpoint{0.000000in}{-0.034722in}}%
\pgfpathcurveto{\pgfqpoint{0.009208in}{-0.034722in}}{\pgfqpoint{0.018041in}{-0.031064in}}{\pgfqpoint{0.024552in}{-0.024552in}}%
\pgfpathcurveto{\pgfqpoint{0.031064in}{-0.018041in}}{\pgfqpoint{0.034722in}{-0.009208in}}{\pgfqpoint{0.034722in}{0.000000in}}%
\pgfpathcurveto{\pgfqpoint{0.034722in}{0.009208in}}{\pgfqpoint{0.031064in}{0.018041in}}{\pgfqpoint{0.024552in}{0.024552in}}%
\pgfpathcurveto{\pgfqpoint{0.018041in}{0.031064in}}{\pgfqpoint{0.009208in}{0.034722in}}{\pgfqpoint{0.000000in}{0.034722in}}%
\pgfpathcurveto{\pgfqpoint{-0.009208in}{0.034722in}}{\pgfqpoint{-0.018041in}{0.031064in}}{\pgfqpoint{-0.024552in}{0.024552in}}%
\pgfpathcurveto{\pgfqpoint{-0.031064in}{0.018041in}}{\pgfqpoint{-0.034722in}{0.009208in}}{\pgfqpoint{-0.034722in}{0.000000in}}%
\pgfpathcurveto{\pgfqpoint{-0.034722in}{-0.009208in}}{\pgfqpoint{-0.031064in}{-0.018041in}}{\pgfqpoint{-0.024552in}{-0.024552in}}%
\pgfpathcurveto{\pgfqpoint{-0.018041in}{-0.031064in}}{\pgfqpoint{-0.009208in}{-0.034722in}}{\pgfqpoint{0.000000in}{-0.034722in}}%
\pgfpathclose%
\pgfusepath{stroke,fill}%
}%
\begin{pgfscope}%
\pgfsys@transformshift{2.313649in}{1.178002in}%
\pgfsys@useobject{currentmarker}{}%
\end{pgfscope}%
\end{pgfscope}%
\begin{pgfscope}%
\definecolor{textcolor}{rgb}{0.000000,0.000000,0.000000}%
\pgfsetstrokecolor{textcolor}%
\pgfsetfillcolor{textcolor}%
\pgftext[x=2.513649in,y=1.139113in,left,base]{\color{textcolor}\rmfamily\fontsize{8.000000}{9.600000}\selectfont 2116 µs}%
\end{pgfscope}%
\begin{pgfscope}%
\pgfsetbuttcap%
\pgfsetmiterjoin%
\definecolor{currentfill}{rgb}{0.000000,0.000000,0.000000}%
\pgfsetfillcolor{currentfill}%
\pgfsetfillopacity{0.000000}%
\pgfsetlinewidth{1.003750pt}%
\definecolor{currentstroke}{rgb}{0.000000,0.000000,0.000000}%
\pgfsetstrokecolor{currentstroke}%
\pgfsetdash{}{0pt}%
\pgfsys@defobject{currentmarker}{\pgfqpoint{-0.034722in}{-0.034722in}}{\pgfqpoint{0.034722in}{0.034722in}}{%
\pgfpathmoveto{\pgfqpoint{0.000000in}{0.034722in}}%
\pgfpathlineto{\pgfqpoint{-0.034722in}{-0.034722in}}%
\pgfpathlineto{\pgfqpoint{0.034722in}{-0.034722in}}%
\pgfpathclose%
\pgfusepath{stroke,fill}%
}%
\begin{pgfscope}%
\pgfsys@transformshift{2.313649in}{1.023113in}%
\pgfsys@useobject{currentmarker}{}%
\end{pgfscope}%
\end{pgfscope}%
\begin{pgfscope}%
\definecolor{textcolor}{rgb}{0.000000,0.000000,0.000000}%
\pgfsetstrokecolor{textcolor}%
\pgfsetfillcolor{textcolor}%
\pgftext[x=2.513649in,y=0.984224in,left,base]{\color{textcolor}\rmfamily\fontsize{8.000000}{9.600000}\selectfont 8244 µs}%
\end{pgfscope}%
\end{pgfpicture}%
\makeatother%
\endgroup%

%% file: deployment.tex
\section{Validation in Field Trial}
\label{sec:deployment}
For a field trial under realistic conditions we equipped five energy harvesting sensor nodes with \sname and deployed them outdoors for environmental sensing of temperature, humidity, air pressure, and particulate matter.
The data \sname provided was used to sustain persistent energy neutral operation by controlling sensing and transmission intervals.
During five weeks the temperatures ranged from \SI{1.5}{\degreeCelsius} (\SI{34.7}{\Fahrenheit}) to \SI{40.1}{\degreeCelsius} (\SI{104.18}{\Fahrenheit}) with intra-day fluctuations of over \SI{32}{\degreeCelsius}~(\SI{57.6}{\Fahrenheit}.).

\begin{figure}
	\centering
	\includegraphics[trim=0 0 0 900,clip, width=0.8\columnwidth]{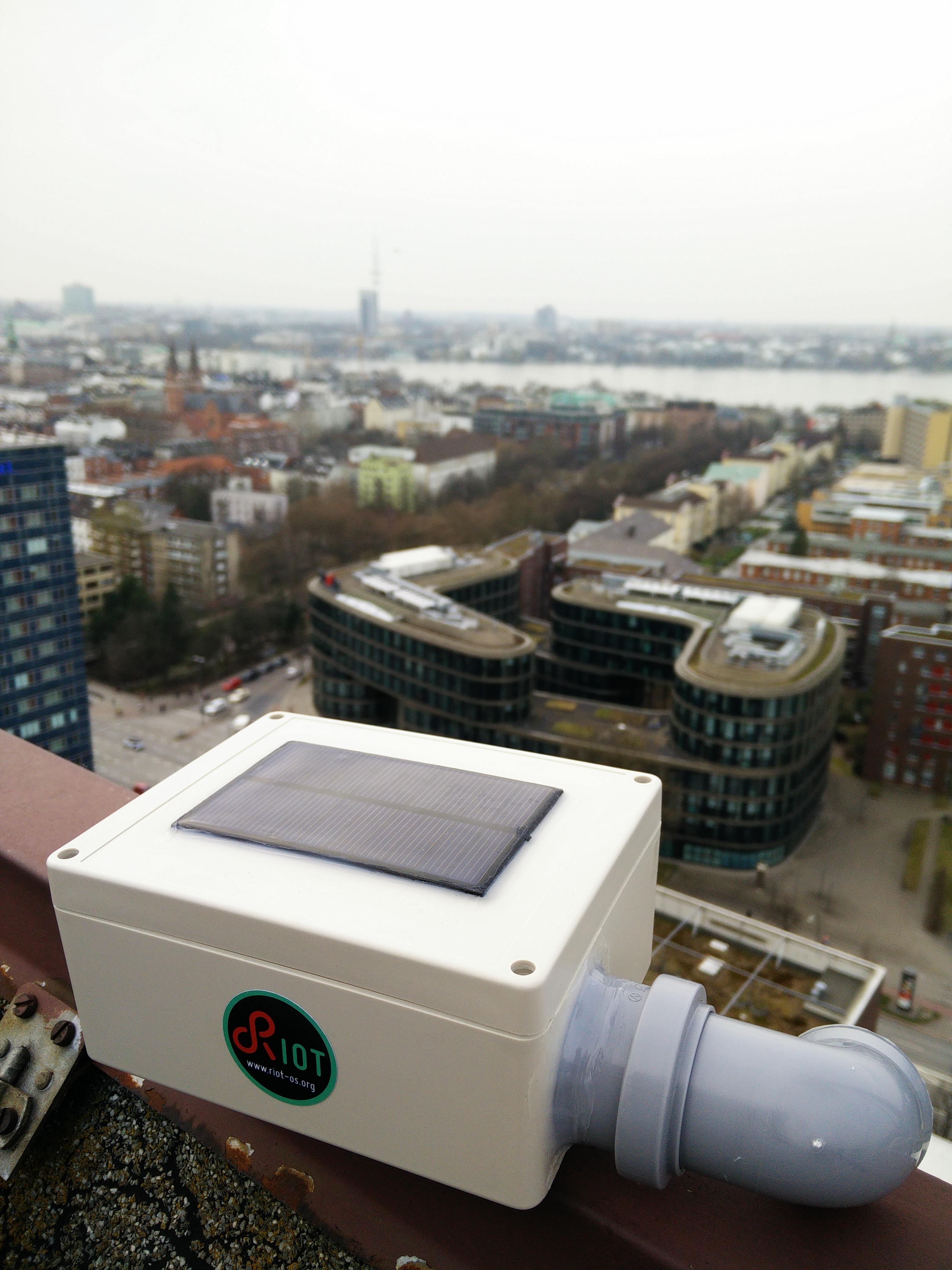}
	\caption{\sname energy-harvesting~system deployed in a watertight box with a tube connector for environmental sensing}
	\label{fig:box_roof_example}
\end{figure}

\paragraph{Deployment Setup}
We deployed five individual \sname-equipped systems on a rooftop directly exposed to the weather, see \autoref{fig:box_roof_example}.
To meet harsh conditions, a waterproof IP65 enclosure box and a tube protected each node and environmental sensors.
We diversified the setting by employing sensors of various manufacturers with different power requirements while using the same generic firmware on all of them.
Energy-harvesting was based on top mounted solar panels.

To connect the IoT devices via IEEE~802.15.4 to the Internet, we used a Raspberry Pi gateway that ran Raspbian Stretch Lite with \texttt{wpan-tools} and \texttt{radvd}.
The IoT network was configured as a pure star topology with node distances from \SIrange{2}{10}{\meter}.

\subsection{Software Logic}

The firmware continuously re-quantifies the charging power and the energy consumed by the task execution, and adapts the duty cycle according to the available energy.The consumption of the application tasks is quantified by the explicit tracing mechanism introduced in \autoref{sec:implementation}.
To determine the charging current, the system issues a measurement cycle with enabled alerting and enters low-power sleep mode immediately.
Once finished, the module wakes up the node, which reads the value.
This procedure is compatible with the lowest possible shutdown modes, which only provide asynchronous wakeup-sources on many MCUs.
Isolated consumption is assessed by disabling the charging circuit to ignore energy provided by the solar panel.
Depending on available and consumed energy, the duty-cycle and intervals for recording fine-grained traces are also adapted.
With the precise and timely knowledge about incoming energy, the node can adapt as early as possible to achieve higher utilization.
The MCU-internal real-time clock (RTC) provided an absolute time reference for the measured data.
A clock drift compensation was implemented to lower the time synchronization interval for the RTC.
Data provided by the simple network time protocol of the RIOT \texttt{sntp} module is therefore used to calibrate the RTC at runtime.
To prevent data loss because of wireless transmission failures, data is logged to the local persistent storage for later investigation.
Sensor data is encoded in a JSON format, transmitted to the gateway via CoAP messages, and finally stored in an external database for post-processing.

\subsection{Results}
\begin{figure}
	\centering
	\input{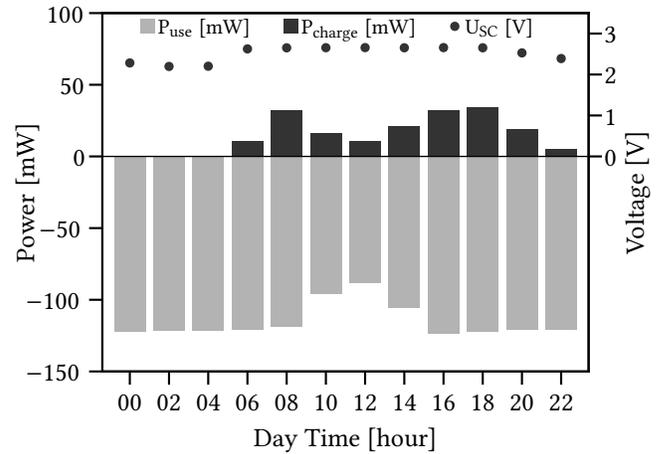}
	\caption{Power consumption and charging over one day observed at one \sname-enabled IoT device (\SI{750}{\milli\ohm} shunt resistor)}
	\label{energy_stats_box_2}
\end{figure}

\autoref{energy_stats_box_2} shows a one-day snapshot of the interplay between power consumption and charging.
Negative bars indicate used power by the node when active, positive bars show charging power when the node is inactive.
Each power value represents an average over a two-hour binning.
The super cap voltage represents the state at the end of each two-hour slot.
A typical sun-cycle can be identified based on the voltage of the super cap as well as the available charging power.
The much high power consumption of the active node shows the need for energy harvesting and duty cycling.

The local memory was utilized to store detailed energy traces of individual task executions for further analysis.
With this onboard data storage, we were able to distinguish erroneous node behavior and failing radio transmissions.
As we also monitored clock drifts, we were able to correlate these drifts with temperature fluctuations.

A lesson learned was that the measurements of the charging power for the inactive nodes need  careful parameterization at high energy availability.
The firmware was configured to limit the duty cycle to a minimum of 10 seconds to keep link stress low for other nodes.
With high energy availability, the time to recharge the super cap after a measurement cycle was sometimes shorter than the selected measurement duration for the charging power.
This effect is visible in \autoref{energy_stats_box_2} at 10:00am to 2:00pm where the power consumption P$_{use}$ drops.
In this case, the charging current was limited because the energy buffer was full and the active period did not consume enough energy to fully utilize availability from the solar panel.
Albeit the measurements still show the average of the actually charged power, it may be of interest to know how much could have been charged with more energy storage capacity.
This allows to proactively schedule more energy before charging starts again, effectively increasing the utilization of the system further.
Additionally, a dynamically switchable load may be helpful in those cases.

For the energy tracing, we used a lower thread priority compared to the rest of the system threads, which proved to be problematic when long-running blocking operations occur.
For example, file system operations sometimes blocked simultaneous tracing.
To circumvent this, either higher priority tracing is required which further increases the invasiveness of the measurement.
Or a non-blocking SD card driver needs to be implemented, which would actually lower the CPU utilization.

\begin{figure}
	\centering
	\input{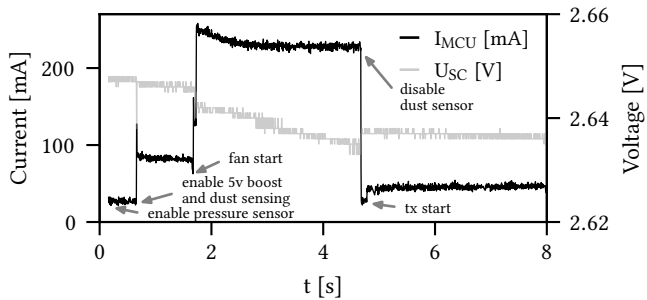}
	\caption{Power trace of a dust sensor execution cycle}
	\label{fig_powertrace}
\end{figure}

A fine-grained trace of the current for a node equipped with a power demanding dust sensor is shown in \autoref{fig_powertrace}.
The consumption of individual sensing steps of the application can be investigated further with this sample.
Shortly after boot, the pressure sensor is powered up and initialized.
Before the dust sensor starts, the boost converter creates a short spike on the power line.
Once the initialization is done, the sensor starts its fan.
It is clearly visible how the startup current slowly settles when the fan reaches its final speed.
Thereafter, the dust sensor is disabled and the collected data is transmitted.

With this kind of self-measurement the node can not only adjust itself to environmental changes but also detect critical component health like raising equivalent series resistance of the super cap or faulty behavior of mechanical components like friction or blocking.

Our field trial showed that \sname is suitable in practice and the collected performance data also underlines that it can be used in even more constrained scenarios \eg with smaller solar panels.
Alternatively, a bigger capacitor could be used to allow higher utilization during nighttime.
With its persistent storage the system could also buffer processing work that is not time critical to perform calculation when energy oversupply occurs.

%% file: conclusions.tex
\section{Conclusions and Outlook}
\label{sec:conclusions}
We presented a hardware-software co-design for self-measuring energy consumption of class-1 IoT nodes \cite{RFC-7228}.
Its generic design is easily portable to many boards due to commodity hardware connected via a standard bus and its seamless integration into a popular IoT OS.
It provides functions to trace code sections on the application level and  threads on the system level.
The module was successfully validated in a comprehensive field trial.

We evaluated the portability and accuracy of our in situ measurement module, as well as its overheads.
Our findings indicate that the solution successfully competes with specialized, highly optimized designs at high accuracy and low overhead. 
The flexibility conjoined with accuracy achieved by Eco helps with adding energy awareness to applications via OS-provided primitives.

In future work, we will focus on providing a generic, light-weight management tool-set.
Extending the support for other types of measurement modules is also considered helpful to make the solution more versatile.
As a rational addition to the OS supported self-measurement, we will work on a separate module that provides default implementations of common energy management algorithms such as EWMA \cite{khzs-pmehs-07}, ENO-MAX \cite{vgb-acdce-07}, and LT-ENO \cite{bsbt-dpmle-14}.